\documentclass[12pt]{article}
\usepackage[utf8]{inputenc}
\usepackage[DIV=13]{typearea}\usepackage{ragged2e}
\usepackage{calligra,amsmath,amsfonts,bbm,mathrsfs,amssymb}
\usepackage{slashed,cancel}
\usepackage{cite}
\usepackage{bm} 
\usepackage{hyperref}
\hypersetup{colorlinks=true,urlcolor=magenta,anchorcolor=blue,citecolor=blue,filecolor=blue,linkcolor=magenta,menucolor=blue, linktocpage=true,pdfproducer=medialab}
\usepackage[english]{babel}
\usepackage{catchfile}
\usepackage{indentfirst}
\usepackage{graphicx}
\usepackage{float}
\usepackage{enumerate}
\usepackage{caption}
\usepackage{subfigure}
\usepackage{multirow}
\usepackage[compat=1.1.0]{tikz-feynman}
\usepackage[normalem]{ulem} 
\usepackage{booktabs}
\usepackage{physics}
\usepackage{url}
\usepackage{soul}
\usepackage{blkarray}
\usepackage{multicol}
\usepackage{longtable}
\usepackage{comment}
\usepackage{pdflscape}
\newcommand{\email}[1]{\href{mailto:#1}{\tt #1}}

\numberwithin{equation}{section}

\newcommand\subsetsim{\mathrel{%
  \ooalign{\raise0.2ex\hbox{$\subset$}\cr\hidewidth\raise-0.8ex\hbox{\scalebox{0.9}{$\sim$}}\hidewidth\cr}}}

\newcommand{\blue}[1]{\color{blue} #1 \color{black}}
\newcommand{\magenta}[1]{\color{magenta} #1 \color{black}}
\newcommand{\be}{\begin{equation}}
\newcommand{\ee}{\end{equation}}
\newcommand{\ba} {\begin{equation}\begin{aligned}}
\newcommand{\ea} {\end{aligned}\end{equation}}

\newcommand{\sL}{\mathscr{L}}

\newcommand{\cO}{\mathcal{O}}

\newcommand{\derp}{\partial}
\newcommand{\hc}{\text{h.c.}}
\newcommand{\nn}{\nonumber}

\newcommand{\ov}[1]{\overline{#1}}
\newcommand{\sign}[1]{\text{sgn}(#1)}

\def\vs{{\textit vs.} }

\def\Tr{{\rm Tr}}

\def\tH{\widetilde{H}}

\newcommand{\eV}{\ \text{eV}}
\newcommand{\keV}{\ \text{keV}}
\newcommand{\TeV}{\ \text{TeV}}
\newcommand{\GeV}{\ \text{GeV}}
\newcommand{\MeV}{\ \text{MeV}}
\newcommand{\fb}{\ \text{fb}}

\newcommand{\PQ}{U(1)_\text{PQ}}
\newcommand{\LN}{U(1)_\text{LN}}

\newcommand{\lrt}[1]{\left(#1\right)}

\newcommand{\lrg}[1]{\left\{#1\right\}}

\DeclareUnicodeCharacter{2212}{-}

\graphicspath{{Plots/}}

\begin{document} 
\renewcommand*{\thefootnote}{\fnsymbol{footnote}}
\begin{titlepage}

\vspace*{-1cm}
\flushleft{\magenta{IFT-UAM/CSIC-23-17}}
\\[1cm]

\begin{center}
\blue{\bf\LARGE\boldmath Visible GeV ALP from}\\[2mm]
\blue{\bf\LARGE\boldmath TeV Vector-like Leptons}\\[4mm]
\unboldmath
\centering
\vskip .3cm
\end{center}
\vskip 0.5  cm
\begin{center}
{\large\bf Arturo de Giorgi}~\footnote{\email{arturo.degiorgi@uam.es}},
{\large\bf Marta Fuentes Zamoro}~\footnote{\email{marta.zamoro@uam.es}}, \\[2mm]
and {\large\bf Luca Merlo}~\footnote{\email{luca.merlo@uam.es}},
\vskip .7cm
{\footnotesize
Departamento de F\'isica Te\'orica and Instituto de F\'isica Te\'orica UAM/CSIC,\\
Universidad Aut\'onoma de Madrid, Cantoblanco, 28049, Madrid, Spain
}
\end{center}
\vskip 2cm
\begin{abstract}
\justify
We present a model where a GeV axion-like-particle (ALP) 
is predicted in a large portion of the parameter space due to the presence of explicit Peccei-Quinn symmetry-breaking terms in an exotic leptonic sector. The latter provides a solution to the muon $g-2$ anomaly, within the framework of the Linear Seesaw neutrino mechanism. The spectrum is extended by a complex scalar singlet only transforming under the Peccei-Quinn symmetry, which generates the ALP. Its couplings with fermions can continuously span over many orders of magnitude, which constitutes a specific feature of this model in contrast to generic ultraviolet constructions. Interestingly, these couplings are suppressed by the ALP characteristic scale that can be as low as the TeV scale, which represents a novel feature of the model and opens up to several phenomenological consequences.
\end{abstract}
\end{titlepage}
\setcounter{footnote}{0}

\pdfbookmark[1]{Table of Contents}{tableofcontents}
\tableofcontents

\renewcommand*{\thefootnote}{\arabic{footnote}}
%
%%%%%%%%%%%%%%%%%%%%%%%%%%%%%%%%%%%%%%%%%%%%%%%%%%%%%%%%%%%
%%%%%%%%%%%%%%%%%%%%%%%%%%%%%%%%%%%%%%%%%%%%%%%%%%%%%%%%%%%
%%%%%%%%%%%%%%%%%%%%%%%%%%%%%%%%%%%%%%%%%%%%%%%%%%%%%%%%%%%
%
\section{Introduction}
\label{sec:intro}

Axions and the more general axion-like-particles (ALPs) are undergoing a revival of interest in these last years. The axion solution to the Strong CP problem consists in the possibility of redefining away the so-called QCD-$\theta$ parameter and, in the original formulations of the late 1970s~\cite{Peccei:1977hh,Weinberg:1977ma,Wilczek:1977pj}, is associated to a global Abelian symmetry - the Peccei-Quinn (PQ) $\PQ$, whose spontaneous breaking gives rise to a Goldstone boson (GB), dubbed as axion, $a$. For this solution to work, the PQ symmetry should be anomalous with QCD and, as a result, a non-perturbative mass $m_a$ for the axion is generated, being inversely proportional to its decay constant $f_a$, such that $m_a\,f_a\sim0.01\GeV^2$. For almost four decades, QCD axion models~\cite{Peccei:1977hh,Weinberg:1977ma,Wilczek:1977pj,Zhitnitsky:1980tq,Dine:1981rt,Kim:1979if,Shifman:1979if} were restricted to a very tiny strip in the $m_a$ \vs $f_a$ parameter space. However, very much recently~\cite{DiLuzio:2016sbl,DiLuzio:2017pfr,Gaillard:2018xgk,Hook:2019qoh,DiLuzio:2020wdo,DiLuzio:2020oah,DiLuzio:2021pxd,
DiLuzio:2021gos,Gavela:2023tzu,Cox:2023dou}, it has been shown that the relation between $m_a$ and $f_a$ can be relaxed, while still solving the Strong CP problem. This represents a turning point in axion physics, providing strong support towards collider searches for not-so-light pseudo-scalars, commonly referred to as ALPs. 

The definition of an ALP in the literature is not unique and here we will consider it as a pseudo-scalar with (dominant) derivative couplings, whose mass is not necessarily associated with any non-perturbative QCD effect. According to the new theoretical results mentioned above, it may or may not be associated with a solution to the Strong CP problem. In the literature, there are many examples of ALPs: associated to flavour dynamics~\cite{Davidson:1981zd,Wilczek:1982rv,Ema:2016ops,Calibbi:2016hwq,Arias-Aragon:2017eww,Arias-Aragon:2022ats,DiLuzio:2023ndz} or neutrino mass generation~\cite{Chikashige:1980qk,Chikashige:1980ui,Gelmini:1980re}; appearing in composite Higgs models~\cite{Merlo:2017sun,Brivio:2017sdm,
Alonso-Gonzalez:2018vpc,Alonso-Gonzalez:2020wst}, supersymmetric contexts~\cite{Bellazzini:2017neg} or in string theories \cite{Witten:1984dg,Choi:2006qj,Svrcek:2006yi,Arvanitaki:2009fg,Cicoli:2012sz}; even as Dark Matter candidates~\cite{Gelmini:1984pe,Berezinsky:1993fm,Lattanzi:2007ux,Bazzocchi:2008fh,Lattanzi:2013uza,Queiroz:2014yna}, or playing a role in cosmological observables~\cite{Ferreira:2018vjj,DEramo:2018vss,Escudero:2019gvw,Arias-Aragon:2020qtn,Arias-Aragon:2020qip, Arias-Aragon:2020shv,Ferreira:2020bpb,Escudero:2021rfi,Araki:2021xdk,DEramo:2021psx,DEramo:2021lgb,DEramo:2022nvb}. Given the existence of so many different contexts in which an ALP arises, part of the community committed itself to the development of an Effective Field Theory (EFT) description encoding the generic ALP features. Following the seminal work in Ref.~\cite{Georgi:1986df}, several studies appeared defining the ALP effective Lagrangian~\cite{Choi:1986zw,Salvio:2013iaa,Brivio:2017ije,Alonso-Alvarez:2018irt,Gavela:2019wzg,Chala:2020wvs,Bonilla:2021ufe,Bauer:2021wjo,Arias-Aragon:2022byr,Arias-Aragon:2022iwl} in order to investigate possible signals at low-energy facilities~\cite{Izaguirre:2016dfi,Merlo:2019anv,Bauer:2019gfk,Bauer:2020jbp,Bauer:2021mvw,Guerrera:2021yss,Gallo:2021ame,Bonilla:2022qgm,Bonilla:2022vtn,deGiorgi:2022vup,Guerrera:2022ykl} and colliders~\cite{Jaeckel:2012yz,Mimasu:2014nea,Jaeckel:2015jla,Alves:2016koo,Knapen:2016moh,Brivio:2017ije,Bauer:2017nlg,Mariotti:2017vtv,Bauer:2017ris,Baldenegro:2018hng,Craig:2018kne,Bauer:2018uxu,Gavela:2019cmq,Haghighat:2020nuh,Wang:2021uyb,deGiorgi:2022oks,Bonilla:2022pxu,Ghebretinsaea:2022djg,Vileta:2022jou, DiLuzio:2024jip}.

In contrast to the QCD axion, whose mass is generated through non-perturbative effects, the ALP mass is typically taken as a free parameter. Only a few studies are present discussing the possible origin of the ALP mass and only identifying the ALP as the Majoron~\cite{Chikashige:1980qk,Chikashige:1980ui,Gelmini:1980re}, that is the Goldstone boson arising from the spontaneous breaking of the Lepton number (LN).
Such a mass may be originated by the explicit breaking of the LN, either via Planck-suppressed operators~\cite{Akhmedov:1992hi,Rothstein:1992rh} or within specific frameworks that deal with the active neutrino mass generation mechanism~\cite{Mohapatra:1982tc,Gu:2010ys,Frigerio:2011in,deGiorgi:2023tvn}. Very recently, Ref.~\cite{deGiorgi:2023tvn} discussed the possibility of giving the Majoron a mass in the range $m_a\subset[1,\,10^5]\keV$ in a specific Type-I Seesaw context, but at the price of a relatively large $f_a$ scale, $f_a\subset[10^{8},\,10^{12}]\GeV$. This is obtained under the assumption of only one source of LN explicit breaking responsible at the same time for the active neutrino masses and the Majoron mass.

\textit{This paper presents a novel framework where an ALP, not being a Majoron, acquires a mass around $\cO(1-10)\GeV$ for a characteristic scale $f_a\in[0.5,\,100]\TeV$ (prior to considering other phenomenological constraints), the latter being much lower than the one presented in Ref.~\cite{deGiorgi:2023tvn}.} Interestingly, the relatively smallness of the ALP decay constant $f_a$ opens up the possibility of a direct search of this exotic state at colliders and flavour facilities. One difference with respect to the context described in Ref.~\cite{deGiorgi:2023tvn} is the presence of an exotic vector-like lepton field transforming as a doublet of $SU(2)_L$ gauge symmetry, owing an explicit PQ symmetry breaking coupling. Although similar results may be obtained with exotic fields transforming under other representations of the electroweak (EW) symmetry, the setup under consideration with an exotic doublet has been recently shown to be extremely appealing, being able to explain the long-standing anomaly in the muon anomalous magnetic moment~\cite{Kannike:2011ng,Dermisek:2013gta,Arcadi:2021cwg, Crivellin:2021rbq,Lu:2021vcp,Guedes:2022cfy}. The $(g-2)_\mu$ is currently at the centre of controversies due to contrasting results in its theoretical prediction with different methods (more on this in Sect.~\ref{sec:Pheno}). Despite this, it offers a challenging benchmark for phenomenology and model-building. 

Refs.~\cite{Arkani-Hamed:2021xlp,Craig:2021ksw,DelleRose:2022ygn} entered into details of a very specific feature of the framework with an additional exotic vector-like lepton field: the dominant (chirally enhanced) contribution to the $(g-2)_\mu$ at $1-$loop arises suppressed by the fourth power of the exotic masses, instead of their square as generically expected. More recently, Ref.~\cite{deGiorgi:2022xhr} explored the possibility of merging the solution to the $(g-2)_\mu$ anomaly with the description of active neutrino masses within the same context, through the Low-Scale Seesaw mechanism~\cite{Kersten:2007vk,Abada:2007ux}. Finally, it is important to note that the non-unitarity of the PMNS matrix induced in this setup leads to possible deviations of the mass of the $W$ gauge boson~\cite{Blennow:2022yfm,Arias-Aragon:2022ats}, as also discussed in Ref.~\cite{deGiorgi:2022xhr} with respect to the recent CDF II result~\cite{CDF:2022hxs}.

Along the lines of Refs.~\cite{Arkani-Hamed:2021xlp,Craig:2021ksw,DelleRose:2022ygn,deGiorgi:2022xhr}, we will consider a simplified one-generation scenario involving only the second lepton generation, adding to the fermion spectrum two right-handed (RH) lepton singlets and a vector-like sibling of the SM lepton EW doublet. We will refer to all the neutral exotic states as Heavy Neutral Leptons (HNLs). From this point of view, it serves as a proof of concept of how a realistic three-generation model could work. Furthermore, besides the SM gauge symmetries, the LN assignments allow to accommodate the Linear Seesaw (LSS) mechanism~\cite{Malinsky:2005bi} and the Lagrangian is made invariant under a PQ symmetry by means of a singlet complex scalar, only transforming under the $\PQ$, whose spontaneous breaking gives rise to the ALP. The exotic fields have an interaction term that explicitly breaks the PQ symmetry, leading to a mass for the ALP induced at $1$-loop, but that does respect the LN at any radiative order. Indeed, the charge assignments under the Abelian symmetries are such that the LN remains exact after the spontaneous and explicit PQ breaking: LN gets only broken explicitly by the traditional terms of the LSS mechanism. On the other hand, for the charged sector, the muon does not have any tree-level mass contribution, being the SM Yukawa interaction forbidden by the PQ symmetry. The muon receives a mass only at $1$-loop and is mediated by the exotic leptons.

The combination of the exotic neutral singlets and lepton doublet leads to the specific suppressed contributions to the $(g-2)_\mu$ already mentioned above, while, on the contrary, neither the ALP nor the radial component of the PQ scalar field contribute to it. A correlation between the contributions to the muon mass and those to the $(g-2)_\mu$ follows, as already pointed out in Ref.~\cite{Arkani-Hamed:2021xlp}. On the other hand, the ALP as well as the HNLs may give signals at colliders and flavour facilities, although a three-generation model would be necessary for a consistent analysis.

As an additional specific feature of the construction presented in this paper, the ALP-muon coupling spans continuously over various orders of magnitude, from $\ll\mathcal{O}(1)/f_a$ to $\mathcal{O}(1)/f_a$. This occurrence is unusual because the naive estimate for the ALP-muon coupling in a given model is typically either $\ll\mathcal{O}(1)/f_a$ or $\mathcal{O}(1)/f_a$, but it cannot continuously vary across different orders of magnitude. Our model serves as a proof-of-concept for the existence of models exhibiting this particular feature. 

In summary, \textit{our model deals with several issues of the SM and theories beyond it: within a single construction, it describes the active neutrino mass scale, provides a solution to the $(g-2)_\mu$ anomaly and predicts an ALP with mass at the GeV scale with a TeV characteristic scale that leads to possibly visible signals at colliders.} Despite this, its description is limited to the second fermion generation and an extension to the three SM families is not unique. As a consequence, while the qualitative results are expected to remain the same in the generalisation, the exact numerical predictions are likely to vary. Nevertheless, our model paves the way for dedicated analyses of these kind of setups. The structure of the paper can be read out in the table of Contents.

%
%%%%%%%%%%%%%%%%%%%%%%%%%%%%%%%%%%%%%%%%%%%%%%%%%%%%%%%%%%%
%%%%%%%%%%%%%%%%%%%%%%%%%%%%%%%%%%%%%%%%%%%%%%%%%%%%%%%%%%%
%%%%%%%%%%%%%%%%%%%%%%%%%%%%%%%%%%%%%%%%%%%%%%%%%%%%%%%%%%%
%
\section{The model description}
\label{sec:Model}

The specific spectrum considered includes the scalar and gauge sector of the SM and its second-generation leptons, assuming that the other fermions are present and with the SM interactions. This guarantees the gauge anomaly cancellation and the consistency of the theory while introducing modifications to the muon and the muonic neutrino, whose phenomenology will be discussed. Besides the left-handed (LH) EW lepton doublet $\ell_L\equiv(\nu_{\mu L},\,\mu_L)^T$ and the right-handed (RH) muon singlet, $\mu_R$, two RH neutrinos, $N_R$ and $S_R$, and a vector-like EW lepton doublet, $\psi$, are introduced. So far, this is the same spectrum considered in Refs.~\cite{Arkani-Hamed:2021xlp,Craig:2021ksw,DelleRose:2022ygn,deGiorgi:2022xhr}. However, we further extend the particle content by adding, on top of the SM Higgs $H$,  a complex scalar field $\phi$, singlet under the SM gauge symmetries.

The whole Lagrangian of the model can be divided into three pieces: the canonical kinetic terms of all the fields, the scalar potential and the mass Lagrangian. The latter is the focus of this section and determines the relevant phenomenology to discuss in this paper, but a few comments on the potential are in order. 
At the renormalisable level, we can distinguish between pure SM couplings, pure $\phi$ ones and mixed interactions:
\ba
V(H,\,\phi)=&\, V_{\text{SM}}(H)+V_{\phi}(\phi)+\lambda_{H\phi}\abs{H}^2\abs{\phi}^2\\
V_{\text{SM}}(H)=&-\mu^2 \left(H^\dagger H\right)+ \lambda \left(H^\dagger H\right)^2\\
V_{\phi}(\phi)=&-\mu_\phi^2 \left(\phi^\ast \phi\right)+ \lambda_\phi \left(\phi^\ast \phi\right)^2\,,
\ea
where $\mu$ and $\mu_\phi$ are masses and $\lambda_i$ are dimensionless coefficients, that need to satisfy the positivity and bounded-from-below conditions. $V_{\text{SM}}(H)$ and $V_{\phi}(\phi)$ are responsible for the spontaneous breaking of the EW and PQ symmetries, while the mixed term is the portal between the SM Higgs and the exotic scalar.

The PQ and EW symmetries get broken by the Higgs and the $\phi$ scalars developing VEVs, at the EW scale $v=246\GeV$ and at $f_a>v$, respectively. We can use the following notation:
\be
H=\dfrac{v+h}{\sqrt2}
\,U\begin{pmatrix}
0\\
1
\end{pmatrix}\,,\qquad\qquad
\phi=\dfrac{f_a+\rho_a}{\sqrt{2}} e^{ia / f_a}\,,
\label{VEVsDef}
\ee
where $h$ is the neutral CP-even Higgs, $U$ the matrix containing the SM GBs, $\rho_a$ the radial component of the singlet $\phi$, and $a$ its angular component to be identified with the ALP. Notice that $f_a$ represents here both the VEV of $\phi$ and the ALP decay constant, guaranteeing canonical kinetic terms for the ALP. We can expand the potential and rewrite the mass term in a compact way as $\frac{1}{2}\Phi^T M_\Phi \Phi$, where $\Phi=\left(h, \rho_a\right)^T$ is the vector of the scalar fields and the mass matrix reads
\be
M_\Phi=\begin{pmatrix}
2\lambda v^2+\dfrac12\lambda_{H\phi}f_a^2& 
\lambda_{H\phi}vf_a\\
\lambda_{H\phi}vf_a& 
2\lambda_{H\phi}f_a^2 +\dfrac12\lambda_{H\phi}v^2
\end{pmatrix}\,.
\ee
If the mixed quartic coupling is of the same order of $\lambda$ or even larger, then the dominant contributions to $M_\Phi$ are the ones proportional to $f_a$. In this case, the physical Higgs and radial mode would have masses of the same order of magnitude and the mixing among the two states would be sizable. The ATLAS collaboration analysed the Higgs signal strengths with $\sqrt{s}=13\TeV$ data with $80\,\fb^{-1}$ of integrated luminosity and obtained a bound on this mixing, assuming a universal suppression on the Higgs couplings, that reads~\cite{ATLAS:2019nkf} 
\be
\sin^2\gamma\lesssim0.09\,\text{at $95\%$ C.L.}\,,
\label{BoundMixingHiggs}
\ee
defining the physical Higgs as
\be
\widehat{h}=\cos{\gamma}\,h-\sin{\gamma}\,\rho_a\,.
\ee
It follows that any sizable mixing is excluded and therefore the mixed quartic coupling has to be necessarily smaller than the pure Higgs quartic one, $\lambda_{H\phi}\ll\lambda$. 

Diagonalising the mass matrix $M_\Phi$, assuming that $f_a> v$ and $\lambda_{H\phi}\ll\lambda$, we can read the masses of the physical Higgs and radial mode,
\be
\widehat{m}_h^2= 2 \lambda v^2+\dfrac{\lambda_{H\phi} f_a^2}{2}\,,\qquad\qquad
\widehat{m}_\rho^2= 2 \lambda_\phi f_a^2+\dfrac{\lambda_{H\phi} v^2}{2}\,,
\label{eq:phys_mass_scalar}
\ee
and the mixing is given by
\be
\tan{\gamma}=\frac{2v\lambda_{H\phi}}{f_a(4\lambda_\phi-\lambda_{H\phi})}\,.
\ee

\begin{figure}[h!]
	\centering
{\includegraphics[width=0.48\linewidth]{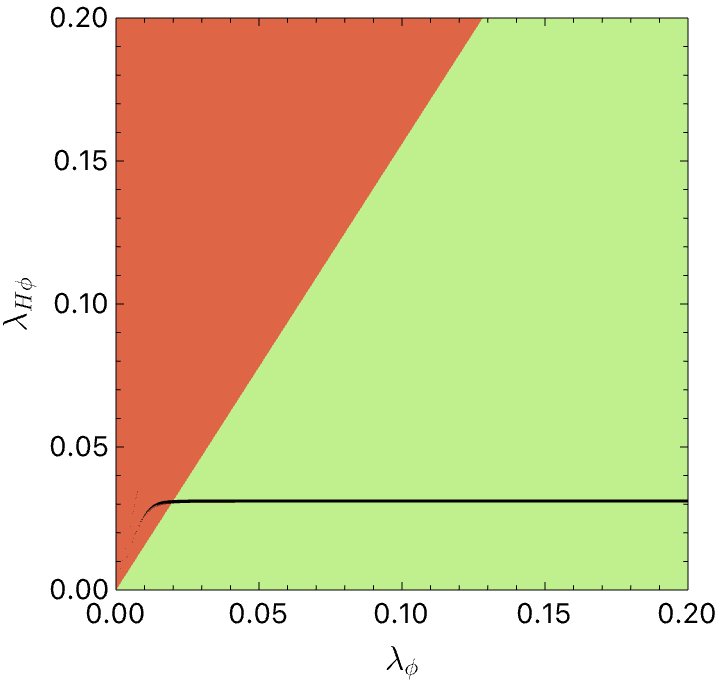}}
     \,
     \caption[]{\em Value of $\sin^2{\gamma}$ as a function of $\lambda_{\phi}$ and $\lambda_{H\phi}$. $f_a=1\TeV$, while $m_a=1 \GeV$. The green (red) region corresponds to values of $\sin^2{\gamma} \leq (>) 0.09$. The black line represents the region of the parameter space in which the measured Higgs mass can be reproduced within its experimental limits and simultaneously $\lambda_{HHH}\in \left(-1.4,6.1\right)\lambda^{SM}_{HHH}$.}
	\label{fig:higgs_int}
\end{figure}

We can then use the bound in Eq.~\eqref{BoundMixingHiggs} to quantify how small the mixed quartic coupling needs to be. We study $\sin^2{\gamma}$ in the parameter space of $\left(\lambda_\phi,\lambda_{H\phi}\right)$ in Fig.~\ref{fig:higgs_int}, the green (red) region corresponding to values of $\sin^2{\gamma} \leq (>)0.09$. We can also impose constraints on the cubic Higgs coupling by using the $\kappa$-formalism~\cite{Peskin:2013xra,Lepage:2014fla,Brod:2013cka}. In this framework,
\be
\kappa_{HHH}=\dfrac{\lambda_{HHH}}{\lambda_{HHH}^\text{SM}}\,,
\ee 
where $\lambda^\text{SM}_{HHH}\equiv m_{h,\text{exp}}^2/2 v^2$, being $m_{h,\text{exp}}=125.20\pm0.11\GeV$~\cite{Workman:2022ynf} the experimental Higgs mass, while $\lambda_{HHH}$ represents the cubic coupling in our model. The most stringent limits on the value of $\kappa_{HHH}$, $-1.4<\kappa_{HHH}<6.1$, have been obtained by the ATLAS collaboration from a study on single- and double- Higgs production~\cite{ATLAS:2022jtk}.  In Fig.~\ref{fig:higgs_int}, the black line represents the region of the parameter space in which the measured Higgs mass can be reproduced within its experimental limits and simultaneously $\lambda_{HHH}\in\left(−1.4, 6.1\right)\lambda_{HHH}^{SM}$. Notice that the numerical results have been obtained  without taking the assumption of $\lambda_{H\phi}\ll \lambda$ (contrary to Eq.\eqref{eq:phys_mass_scalar}). It can be seen that $\lambda_{H\phi}\sim 0.03$ at most.

Besides the contribution from the mixed quartic coupling, the Higgs parameters receive loop corrections not only from the SM fields but also from the new exotic fermions that, being the heaviest, will provide the largest contributions. This is nothing more than the well-known Higgs Hierarchy problem that we will not attempt to solve here. 

To sum up, in what follows, without loss of generality, but avoiding unnecessary complications in the computation of the relevant phenomenology of this model, we will simply assume that i) the Higgs and the $\phi$ scalars develop VEVs $f_a>v$, ii) the mixed quartic coupling is small enough to consider decoupled the pure Higgs and $\phi$ sectors, and iii) the radiative corrections to the pure Higgs sector are simply reabsorbed in the bare Lagrangian parameters. 
Moreover, as a further hypothesis, we will assume that iv) the radial component $\rho_a$ achieves a sufficiently large mass to disentangle it from the low-energy Lagrangian. We will comment, however, on how it would affect the main aspects of our phenomenological analysis, showing that it would not have any relevant impact on the $(g-2)_\mu$, even if relatively light. \\

Regarding the part of the Lagrangian responsible for giving masses to the second SM lepton generation and the exotic leptons, we will assume the following expression
\begin{equation}
	\begin{aligned}
		 -\sL_Y=&\phantom{+}Y_N\ov{\ell_L} \tH N_R+Y_R \ov{\psi_L} H \mu_R+\\
		&+\delta_{x,0} \Lambda \overline{N_R^c} S_R+\delta_{|x|,1} \alpha_N \phi^{(\ast)} \overline{N_R^c} S_R+\delta_{y,0} M_\psi \ov{\psi_L} \psi_R+\delta_{|y|,1} \alpha_\psi\phi^{(\ast)} \ov{\psi_L} \psi_R+ \\
		& +Y_V \overline{S_R^c} \tH^{\dagger} \psi_R+Y_{V^\prime} \ov{\psi_L} \tH N_R +\epsilon Y_S\ov{\ell_L} \tH  S_R+\hc\,, \\
	\end{aligned}
	\label{lmodel}
\end{equation}
where $\tH\equiv i \sigma_2 H$, being $\sigma_2$ the second Pauli matrix, and $\delta_{i,j}$ is the Kronecker delta. The quantities $Y_N$, $Y_R$, $Y_V$, $Y_{V'}$, $Y_S$, $\alpha_N$ and $\alpha_\psi$ are dimensionless parameters and $\Lambda$ and $M_\psi$ are instead masses. On the other hand, $x$ and $y$ are variables that can only acquire three values, either $0$ or $\pm1$, distinguishing in this way different realisations. For example, the two terms containing the bilinear $\overline{N_R^c} S_R$ cannot be present simultaneously: the idea is that for symmetry reasons this bilinear appears proportional either to the scale $\Lambda$ or to the product $\alpha_N \phi$ or $\alpha_N \phi^\ast$, corresponding respectively to $x=0,\,1,\,-1$. Similarly for $y$ and the $\ov{\psi_L} \psi_R$-term. Finally, $\epsilon$ is a small parameter and is responsible for the active neutrino mass {\it \`a la} LSS mechanism. 

The Lagrangian described above, neglecting the last line of Eq.~\eqref{lmodel}, shows the following symmetries,
\be
\Big[SU(2)_L\times U(1)_{Y}\Big]_\text{gauge}\times\Big[\LN\times \PQ\Big]_\text{global}\,,
\ee
and the transformation properties of the fields are listed in Tab.~\ref{table:particlespectrum}. Notice that the invariance under the PQ symmetry implies that only two terms in the second line of Eq.~\eqref{lmodel} are allowed, depending on the explicit charges of the fields: one among the terms proportional to $x$, and one among those proportional to $y$.

\begin{table}[ht!]
	\centering
	\begin{tabular}{c||c|c||c|c|}
		&$SU(2)_{L}$&$U(1)_{Y}$ &$U(1)_{L}$ & $\PQ$\\
		\hline \hline
		$\ell_L$&$\bf{2}$&$-1/2$&$1$&$n_{N_R}$\\
		$\mu_{R}$&$1$&$-1$&$1$&$n_{\psi_L}$\\
		$H$&$\bf{2}$&$1/2$&$0$&$0$\\
		\hline
		$N_{R}$&$1$&$0$&$1$&$n_{N_R}$\\
		$S_{R}$&$1$&$0$&$-1$&$n_{S_R}$\\		
		\hline	
    		$\psi_{L}$&$\bf{2}$&$-1/2$&$1$&$n_{\psi_L}$\\
		$\psi_{R}$&$\bf{2}$&$-1/2$&$1$&$n_{\psi_R}$\\
		\hline
		$\phi$&$1$&$0$&$0$&$n_\phi$\\
		\hline		
	\end{tabular}
	\caption{\em Transformation properties of the SM leptons $\ell_{L}$ and $\mu_{R}$, the Higgs doublet H, the HNLs $N_R$ and $S_R$, the vector-like EW lepton doublet $\psi$ and the scalar $\phi$ under $SU(2)_L\times U(1)_{Y}\times\LN\times \PQ$. The PQ charges of $\ell_L$ and $\mu_R$ have been written in terms of the others according to the first line of the Lagrangian in Eq.~\eqref{lmodel}.}
	\label{table:particlespectrum}
\end{table}

The PQ charges of $S_R$ and $\psi_R$ are not arbitrary but can be written in terms of the other charges after fixing the values of the variables $x$, and $y$:
\be
\begin{cases}
x=0 \quad &\Longrightarrow\quad n_{S_R}=-n_{N_R}\\[2mm]
x=1 \quad &\Longrightarrow\quad n_{S_R}=-n_{N_R}- n_\phi\\[2mm]
x=-1 \quad &\Longrightarrow\quad n_{S_R}=-n_{N_R}+ n_\phi\\[2mm]
\hline
&\\[-5mm]
y=0  \quad &\Longrightarrow\quad n_{\psi_R}=n_{\psi_L}\\[2mm]
y=1  \quad &\Longrightarrow\quad n_{\psi_R}=n_{\psi_L}- n_{\phi} \\[2mm]
y=-1\quad &\Longrightarrow\quad n_{\psi_R}=n_{\psi_L}+ n_{\phi} 
\end{cases}
\ee

The different combinations of possible values for $n_{S_R}$ and $n_{\psi_R}$ give rise to nine different models, that however do not lead to nine distinct phenomenologies. In particular, one of the realisations matches the one discussed in Ref.~\cite{deGiorgi:2022xhr}, that is the option with $x=0=y$, corresponding to the case with the scalar $\phi$ not coupling to any fermion. In this case, the interactions proportional to $Y_{V}$ and $Y_{V'}$ have to be allowed in the Lagrangian to solve the $(g-2)_\mu$ anomaly, and this leads to $n_{N_R}=n_{\psi_R}$. As a result, the two Abelian symmetries coincide and the only explicit breaking is the one associated with $\epsilon Y_S$ that provides a mass to the active neutrino. In the end, the ALP would not couple to any particle and therefore it would be secluded in a dark sector, without taking part in the visible phenomenology. Notice that in this case the traditional tree-level muon Yukawa can be written in the Lagrangian as well as a tree-level mixing between the SM lepton doublet and the exotic one, that is $M_L\ov{\ell_L}\psi_R$. These tree-level terms have been considered in Ref.~\cite{deGiorgi:2022xhr}, where the impact of the arbitrary large-scale $M_L$ has been studied. 

Focusing instead on the cases where the scalar singlet, with PQ charge $n_\phi\neq0$, does couple to the leptons, we can identify four models with distinct phenomenological features, differing only in the type of terms in the second line of Eq.~\eqref{lmodel}. All in all, we can characterise them in terms of the PQ charges of $n_{S_R}$ and $n_{\psi_R}$.
\begin{description}
\item[Model A:] The values of the variables are $(x,y)=(\pm1,0)$, such that $n_{\psi_R}=n_{\psi_L}$ and there are two possible choices for the PQ charge assignment of $S_R$, $n_{S_R}=-n_{N_R}\mp n_\phi$, corresponding to the following Lagrangian,
\begin{equation}
\begin{aligned}
-\sL_Y^\text{A}=&\phantom{+}Y_N\ov{\ell_L} \tH N_R+Y_R \ov{\psi_L} H \mu_R
+\alpha_N \phi^{(\ast)} \overline{N_R^c} S_R+M_\psi\ov{\psi_L} \psi_R+\\
&+Y_V \ov{S_R^c}\tH^\dagger\psi_R+Y_V^\prime \ov{\psi_L}\tH N_R+\epsilon Y_S\ov{\ell_L} \tH S_R+\hc\,.
\end{aligned}
\label{LagA}
\end{equation}

\item[Model B:] The values of the variables are $(x,y)=(0,\pm1)$, corresponding to $n_{S_R}=-n_{N_R}$ and two alternatives are possible for the PQ charge assignment of $\psi_R$, $n_{\psi_R}=n_{\psi_L}\mp n_\phi$. The resulting Lagrangian reads
\begin{equation}
\begin{aligned}
-\sL_Y^\text{B}=&\phantom{+}Y_N\ov{\ell_L} \tH N_R+Y_R \ov{\psi_L} H \mu_R
+\Lambda \overline{N_R^c} S_R+\alpha_\psi\phi^{(\ast)}\ov{\psi_L} \psi_R+\\
&+Y_V \ov{S_R^c}\tH^\dagger\psi_R+Y_V^\prime \ov{\psi_L}\tH N_R+\epsilon Y_S\ov{\ell_L} \tH S_R+\hc\,.
\end{aligned}
\label{LagB}
\end{equation}

\item[Model C:] The values of the variables are $(x,y)=(1,1)$ or $(x,y)=(-1,-1)$, such that there are no direct masses, but instead Yukawa-like interactions with the insertion of either only $\phi$ in the two terms or only $\phi^\ast$. The associate Lagrangian is given by
\begin{equation}
\begin{aligned}
-\sL_Y^\text{C}=&\phantom{+}Y_N\ov{\ell_L} \tH N_R+Y_R \ov{\psi_L} H \mu_R
+\alpha_N \phi^{(\ast)} \overline{N_R^c} S_R+\alpha_\psi\phi^{(\ast)}\ov{\psi_L} \psi_R+\\
&+Y_V \ov{S_R^c}\tH^\dagger\psi_R+Y_V^\prime \ov{\psi_L}\tH N_R+\epsilon Y_S\ov{\ell_L} \tH S_R+\hc\,.
\end{aligned}
\label{LagC}
\end{equation}

\item[Model D:] This realisation is very similar to the previous one, but with either $(x,y)=(1,-1)$ or  $(x,y)=(-1,1)$, corresponding to Yukawa-like terms obtained inserting $\phi$ and $\phi^\ast$, such that both of them are simultaneously present in the Lagrangian. The Lagrangian associated to $(x,y)=(1,-1)$ is then
\begin{equation}
\begin{aligned}
-\sL_Y^\text{D}=&\phantom{+}Y_N\ov{\ell_L} \tH N_R+Y_R \ov{\psi_L} H \mu_R
+\alpha_N \phi \overline{N_R^c} S_R+\alpha_\psi\phi^\ast\ov{\psi_L} \psi_R+\\
&+Y_V \ov{S_R^c}\tH^\dagger\psi_R+Y_V^\prime \ov{\psi_L}\tH N_R+\epsilon Y_S\ov{\ell_L} \tH S_R+\hc\,,
\end{aligned}
\label{LagD}
\end{equation}
while the one associated to $(x,y)=(-1,1)$ is obtained simply interchanging $\phi$ with $\phi^\ast$. We will see that this difference with respect to Model C has a deep impact on the corresponding phenomenology.
\end{description}

While we determined the symmetries and the Abelian charge assignments looking only at the terms in the first line of the Lagrangian densities above, we still have to discuss the impact of the terms proportional to $Y_V$, $Y_{V'}$ and $Y_S$. The first two are trivially invariant under LN, while the last one represents an explicit LN breaking. We thus expect a mass for the active neutrino proportional to $\epsilon Y_S$.\footnote{Alternatively, one could keep a more general approach, without fixing the LN charges equal to $\pm1$. Doing so, it would turn out that a neutrino mass is originated only if the terms proportional to $Y_N$ and $\epsilon Y_S$ are both non-vanishing.}

On the other hand, the terms proportional to $Y_V$, $Y_{V'}$ and $Y_S$ represent possible sources of explicit PQ symmetry breaking, depending on the specific choice of the PQ charges. It is interesting to further investigate the conditions that lead to this breaking as they would directly tell whether a mass for the ALP is expected or not without performing any explicit computation. We can use the technique of the spurions to make these terms formally PQ invariant: it consists in promoting $Y_V$, $Y_{V'}$ and $Y_S$ to non-dynamical fields with transformation properties under $\PQ$, such that  
\be
n_{Y_V}=-n_{\psi_R}-n_{S_R}\,,\qquad\qquad
n_{Y_{V'}}=n_{\psi_L}-n_{N_R}\,,\qquad\qquad
n_{Y_S}=n_{N_R}-n_{S_R}\,.
\ee
For each model, {\it if there is a PQ charge assignment such that $n_{Y_V},n_{Y_{V'}},n_{Y_S}=0$, then the corresponding term is not explicitly breaking $\PQ$}. This is easy to compute considering the conditions on $n_{S_R}$ and $n_{\psi_R}$ that characterise each construction. We find:
\be
\begin{cases}
\text{Model A:}\quad\Longrightarrow\quad
n_{Y_V}=-n_{Y_{V'}}\pm n_\phi\qquad
&n_{Y_S}=2n_{N_R}\pm n_\phi\\[2mm]
\text{Model B:}\quad\Longrightarrow\quad
n_{Y_V}=-n_{Y_{V'}}\pm n_\phi\qquad
&n_{Y_S}=2n_{N_R}\\[2mm]
\text{Model C:}\quad\Longrightarrow\quad
n_{Y_V}=-n_{Y_{V'}}\pm 2 n_\phi\qquad
&n_{Y_S}=2n_{N_R}\pm n_\phi\\[2mm]
\text{Model D:}\quad\Longrightarrow\quad
n_{Y_V}=-n_{Y_{V'}}\qquad
&n_{Y_S}=2n_{N_R}\pm n_\phi\,,
\end{cases}
\label{PatternSymBreaking}
\ee
where the signs in front of $n_\phi$ are not arbitrary and correspond to the specific Lagrangian densities in Eqs.~\eqref{LagA}--\eqref{LagD}.

First of all, for Models A, C and D, there is the possibility to fix $n_{N_R}=\mp n_\phi/2\neq0$ and then $n_{Y_S}=0$. For Model B, this same result can also be achieved, but for $n_{N_R}=0=n_{S_R}$. As a result, for all the models, the term proportional to $Y_S$ never explicitly breaks the PQ symmetry and therefore it does not lead to any contribution to the ALP mass.

The discussion on $n_{Y_V}$ and $n_{Y_{V'}}$ is slightly more involved. Starting for simplicity with Model D, choosing $n_{\psi_L}=n_{N_R}$ then $n_{Y_V}=0=n_{Y_{V'}}$ and thus there is no source at all of PQ explicit breaking and the ALP would remain exactly massless at any loop order. \\
For Models A, B, and C, we have to distinguish between having both $Y_V$ and $Y_{V'}$ different from zero and when one and only one of the two is vanishing. In the latter case, it is always possible to find a PQ charge assignment such that the spurion charge of the surviving term is zero: it follows that the only would-be-PQ explicit breaking term actually preserves the symmetry and thus the ALP remains massless. On the contrary, when $Y_V\neq0\neq Y_{V'}$, it is never possible to have $n_{Y_V}=0=n_{Y_{V'}}$ at the same time and therefore the ALP would acquire a mass. This condition implies that a massive ALP only arises if $n_{\psi_R}\neq -n_{S_R}$ and $n_{\psi_L}\neq n_{N_R}$.

{\it All in all, we may expect a massive ALP only in Models A, B and C, whose mass is necessarily proportional to the product of the two parameters $Y_V\, Y_{V'}$ and the two mass scales of the HNLs, i.e.}
\begin{equation}
\label{eq:ALP-mass-short}
    m_a^2 \propto Y_V Y_{V'}\Lambda M_\psi\,,
\end{equation}
where $\Lambda$ and $M_\psi$ can be dynamically generated, i.e. $\Lambda=\alpha_N f_a/\sqrt2$ and $M_\psi=\alpha_\psi f_a/\sqrt2$.

%
%%%%%%%%%%%%%%%%%%%%%%%%%%%%%%%%%%%%%%%%%%%%%%%%%%%%%%%%%%%
%
\subsection{Lepton mass Lagrangian}

Before moving to the determination of the ALP mass and interactions, we discuss the lepton masses following a very similar notation to Ref.~\cite{deGiorgi:2022xhr}. Once the Higgs and the scalar singlet, always present in the four considered models, get a VEV, the mass Lagrangian for the neutral and charged sectors reads 
\be
\begin{aligned}
    -\sL_Y\supset \frac{1}{2}\ov{\chi}\mathcal{M}_\chi \chi^c+ \ov{\zeta_L}\mathcal{M}_\zeta \zeta_R+\hc\,,
\end{aligned}
\ee
where we defined the neutral lepton multiplet $\chi$ and the charged one $\zeta$ as
\begin{equation}
    \chi\equiv (\nu_L, N_R^c,S_R^c,\psi_L^0,\psi_R^{0\,c})^T,\qquad \zeta\equiv (\mu,\psi^-)^T\,,
    \label{multiplet}
\end{equation}
and the charge conjugation operation is defined by $N^c_R \equiv \mathcal{C} \ov{N_R}^T$ with $\mathcal{C} $ being the charge conjugation matrix.
The mass matrices $\mathcal{M}_\chi$ and $\mathcal{M}_\zeta$ read
\begin{equation}
\label{eq:massmatrix}
    \mathcal{M}_{\chi}=\left(\begin{array}{ccccc}
		0 & m_{N} & \epsilon m_{S} & 0 & 0 \\
		m_{N} & 0 & \Lambda & m_{V^{\prime}} & 0 \\
		\epsilon m_{S} & \Lambda & 0 & 0 & m_{V} \\
		0 & m_{V^{\prime}} & 0 & 0 & M_{\psi} \\
		0 & 0 & m_{V} & M_{\psi} & 0
	\end{array}\right), \quad 
    \mathcal{M}_\zeta=\left(\begin{array}{cc}
        0 & 0 \\
         m_R&M_\psi 
    \end{array}\right)
    \,,
\end{equation}
where $m_X\equiv Y_X v/\sqrt{2}$ and, with an abuse of notation, $M_\psi$ stands either for the direct Lagrangian mass $M_\psi$ in model A or for $\alpha_\psi f_a /\sqrt{2}$ in models B, C, and D, and similarly for  $\Lambda$ in the corresponding models.

The neutral mass matrix $\mathcal{M}_{\chi}$ is very similar to the one in Ref.~\cite{deGiorgi:2022xhr}: they coincide in the limit of vanishing $M_L$ and $\mu$ terms, that are the direct mass term $M_L\ov{\ell_L}\psi_R$ already discussed previously and the traditional explicit Lepton number breaking term usually associated to the Inverse Seesaw mechanism, $\mu \ov{S^c_R}S_R$. The mass eigenvalues and eigenvectors can simply be taken from that reference fixing $M_L=0$ and $\mu=0$:
\ba
\widehat{m}_\nu &=-\dfrac{2\,\epsilon\,m_N\, m_S}{\Lambda} \,,\\[2mm]
\widehat{m}_{N_R} &= \Lambda + \dfrac{m_N^2}{2 \Lambda}+ \dfrac{1}{4}\left[\dfrac{(m_V+m_{V'})^2}{\Lambda-M_\psi}+\dfrac{(m_V-m_{V'})^2}{\Lambda+M_\psi}\right]\,,\\[2mm]
\widehat{m}_{S_R}&= \Lambda+ \dfrac{m_N^2}{2 \Lambda}+ \dfrac{1}{4}\left[\dfrac{(m_V+m_{V'})^2}{\Lambda-M_\psi}+\dfrac{(m_V-m_{V'})^2}{\Lambda+M_\psi}\right]\,,\\[2mm]
\widehat{m}_{\psi^0} &= M_\psi -\dfrac{1}{4}\left[\dfrac{(m_V+m_{V'})^2}{\Lambda-M_\psi}-\dfrac{(m_V-m_{V'})^2}{\Lambda+M_\psi}\right]\,,
\label{FinalMassesNeutral}
\ea
and
\be
\widehat{\chi}:
\left\{\begin{array}{l}
\widehat\nu_L = \nu_L-\dfrac{m_N}{\Lambda}\,S^c_R\,,\\[3mm]
\widehat N_R = \dfrac{N_R+S_R}{\sqrt2} +\dfrac{m_N}{\sqrt{2}\Lambda}\nu_L^c 
+\dfrac{1}{2}\left[\dfrac{m_V+m_{V'}}{\Lambda-M_\psi} \dfrac{\psi^{0c}_L+\psi^0_R}{\sqrt2}-
\dfrac{m_V-m_{V'}}{\Lambda+M_\psi}\dfrac{\psi^{0c}_L-\psi^0_R}{\sqrt2}\right] \,,\\[3mm]
\widehat S_R = i\lrg{-\dfrac{N_R-S_R}{\sqrt2}+\dfrac{m_N}{\sqrt{2}\Lambda}\nu_L^c-\dfrac{1}{2}\left[\dfrac{m_V-m_{V'}}{\Lambda+M_\psi} \dfrac{\psi^{0c}_L+\psi^0_R}{\sqrt2}-
\dfrac{m_V+m_{V'}}{\Lambda-M_\psi}
\dfrac{\psi^{0c}_L-\psi^0_R}{\sqrt2}\right]}  \,,\\[3mm]
\widehat\psi^0_L = \dfrac{\psi^0_L+\psi^{0c}_R}{\sqrt2} -
\dfrac{1}{2} \left[\dfrac{m_V+m_{V'}}{\Lambda-M_\psi} \dfrac{N^c_R+S^c_R}{\sqrt2}+\dfrac{m_V-m_{V'}}{\Lambda+M_\psi}\dfrac{N^c_R-S^c_R}{\sqrt2}\right]\,,\\[3mm]
\widehat\psi^0_R = i\lrg{-\dfrac{\psi^{0c}_L-\psi^0_R}{\sqrt2}-\dfrac{1}{2} \left[\dfrac{m_V-m_{V'}}{\Lambda+M_\psi} \dfrac{N_R+S_R}{\sqrt2}+\dfrac{m_V+m_{V'}}{\Lambda-M_\psi}\dfrac{N_R-S_R}{\sqrt2}\right]}\,,
\end{array}\right.
\label{ChiMassEig}
\ee
In the previous expressions, we neglected subleading terms that are proportional to \linebreak $(\epsilon m_S)/(\Lambda,M_\psi)$  and to $\lrt{v/(\Lambda,\,M_\psi)}^2$ for the masses and to $v/(\Lambda,\,M_\psi)$ for the eigenstates. In the numerical analysis, we will exclude the parameter space where these terms are larger than the $20\%$ of the shown expressions. 

The charged sector presents a completely different texture  as both entries in the first row of $\mathcal{M}_\zeta$ are vanishing. The diagonalisation of the mass matrix leads to the following eigenvalues
\begin{equation}
    \widehat{m}_\mu=0,\qquad\qquad 
    \widehat{m}_{\psi^-}=\sqrt{M_\psi^2+m_R^2}\approx M_\psi\left[1+\frac{1}{2}\left(\frac{m_R}{M_\psi}\right)^2\right]\,,
\end{equation}
where terms that have a relative suppression higher than $(v/(\Lambda,M_\psi))^2$ have been neglected. Interestingly, only a rotation of the right-handed fields is needed, so that the mass eigenstates at tree-level read
\begin{equation}
\label{eq:charged-mixing}
    \widehat{\zeta}: 
    \left\{\begin{array}{ll}
        \widehat{\mu}_L=\mu_L\,,\qquad\qquad &  \widehat{\mu}_R=\dfrac{M_\psi}{\sqrt{M_\psi^2+m_R^2}}\mu_R-\dfrac{m_R}{\sqrt{M_\psi^2+m_R^2}}\psi^-_R\,,  \\[2mm]
        \widehat{\psi}^-_L=\psi^-_L\,,& \widehat{\psi^-_R}=\dfrac{m_R}{\sqrt{M_\psi^2+m_R^2}}\mu_R+\dfrac{M_\psi}{\sqrt{M_\psi^2+m_R^2}}\psi^-_R\,,
    \end{array}\right.
\end{equation}
The muon mass is vanishing at tree-level, but we do expect radiative contributions as the global symmetry protecting the muon mass term is explicitly broken. Loop computations have been performed with the help of the \textsc{Mathematica} packages \textsc{Package-X} and \textsc{FeynCalc}~\cite{ Patel:2016fam, Shtabovenko:2020gxv}. The direct computation of the 1-loop diagrams in the $\ov{\text{MS}}$-scheme contributing to the muon bilinear $\ov{\mu}_L\mu_R$ gives 
\be
\delta m_\mu=-\dfrac{m_N\,m_R\,\Lambda}{8\,\pi^2\,v^2}
    \left(\dfrac{m_V}{M_\psi}+\dfrac{m_{V'}}{\Lambda}\right)
    \left[1+\dfrac{1}{M_\psi^2-\Lambda^2}\left(M^2_\psi
    \log{\dfrac{\mu_\text{R}^2}{M_\psi^2}}-
    \Lambda^2
    \log{\dfrac{\mu_\text{R}^2}{\Lambda^2}}\right)\right]\,,
\label{deltammu}
\ee
where we neglected $\mathcal{O}(\Lambda^{-2},M_{\psi}^{-2})$ corrections.
As $\Lambda$ and $M_\psi$ are of the same orders of magnitude, we do not expect significant dependence on the renormalisation point and thus we can fix $\mu_\text{R}=\max\{\Lambda,\,M_\psi\}$. 

This is not, however, the end of the story. Radiative contributions are also present in the other entries of the charged lepton mass matrix and in the kinetic terms, inducing 1-loop mixing of the $\mu$ and $\psi^-$ self energies
\begin{equation}
\label{eq:self-energy-mix}
    \Sigma(\slashed{p})\equiv \begin{pmatrix}
        (1+\Sigma^{p,\mu\mu})\slashed{p}-\delta m_\mu &  \Sigma^{p,\mu\psi}\slashed{p}-(\delta m_R^\dagger P_L+\delta M_L P_R)\\[2mm]
         \Sigma^{p,\psi\mu}\slashed{p}-(\delta M_L^\dagger P_L+\delta m_R P_R) &  (1+\Sigma^{p,\psi\psi})\slashed{p}-(M_\psi+\delta M_\psi)
    \end{pmatrix}\,,
\end{equation}
where
\begin{equation}
    \Sigma^{p,ij}\equiv \Sigma_L^{p,ij}P_L+\Sigma_R^{p,ij}P_R\,,
\end{equation}
and $\delta m_{\mu,R},\,\delta M_{\psi,L}$ are the $\slashed{p}$-dependent and independent contributions of the self-energies, respectively.  
%The full structure of the self-energy matrix can be found in App.~\ref{appendix:Charged-mass-1-loop}. 
Such a matrix can be diagonalised by a non-unitary rotation and the poles of its inverse identify the pole-masses of the $\mu$ and  $\psi^-$ fields. The mixing between the left-handed leptons reads
\begin{equation}
\label{eq:left-mixing}
    \begin{cases}
        \widehat{\mu}_L=\mu_L\left(1+\dfrac{\Sigma^{p,\mu\mu}_L}{2}\right)+\psi_L^-\left(-\dfrac{\delta M_L}{M_\psi}+\Sigma_L^{p,\mu\psi}\right)\,,\\[2mm]
         \widehat{\psi}^-_L=\dfrac{\delta M_L}{M_\psi}\mu_L+\psi_L^-\left(1+\dfrac{\Sigma^{p,\psi\psi}_L}{2}\right)\,,
    \end{cases}
\end{equation}
where we neglected $\mathcal{O}(1/M_{\psi}^2)$ and $\mathcal{O}(\text{2-loops})$ corrections. We can deduce that the renormalisation of the kinetic terms, proportional to $\Sigma^{p,\mu\mu}$, would lead to only subleading effects to the muon mass term and therefore can be neglected. In fact, considering NP contributions to the self-energies, we find that for $\mu\sim \Lambda\sim M_\psi$
\begin{equation}
    \Sigma_{L,R}^{p,\mu\mu}\sim \dfrac{1}{16\pi^2}\log\left(\dfrac{\Lambda^2}{M_W^2}\right)\sim 3\%\,.
\end{equation}
As the wave function renormalisation corrects the pole mass prediction with a factor $\sim 1+\Sigma$, its impact on the 1-loop direct mass term is negligible. Thus, the muon pole-mass at 1-loop is then simply given by 
\begin{equation}
\widehat{m}_\mu=\delta m_\mu
\end{equation}
as defined in Eq.~\eqref{deltammu}. Notice that for $\Lambda=M_\psi$, the expression in the squared bracket of Eq.~\eqref{deltammu} is vanishing and the 2-loop computation would be necessary. Moreover, whenever $\Lambda\approx M_\psi$, a realistic value for the muon mass would require large values for the Yukawa couplings to compensate for the partial cancellation of the terms in the squared bracket. All in all, the muon mass prediction is delicate whenever $\Lambda\approx M_\psi$. We will come back to this aspect in the phenomenological analysis. 

To conclude, the second eigenvalue also receives radiative corrections, both proportional to $\delta M_L$ and $\Sigma^p_L$. However, these corrections are very much suppressed with respect to the leading tree-level contribution and we will neglect them in the numerical analysis that follows. We only report the result for $\delta M_L$ as it will enter the ALP couplings: in the $\ov{\text{MS}}$-scheme, it reads
\be
\delta M_L= \dfrac{m_N m_V \Lambda}{8\pi^2 v^2}\left[1+\log\left(\frac{\mu_R^2}{\Lambda^2}\right)\right]\,.
\label{deltaML}
\ee
As for $\delta m_\mu$, in the numerical analysis we will fix $\mu_\text{R}=\max\{\Lambda,\,M_\psi\}$.

%%%%%%%%%%%%%%%%%%%%%%%%%%%%%%%%%%%%%%%%%%%%%%%%%%%%%
\subsection{ALP Lagrangian}

In this section, we analyse the ALP Lagrangian and the interactions of the ALP, in particular with SM fields.
We will mainly work in the so-called \textit{chirality-flipping} basis, that is the one obtained from the Lagrangian in Eq.~\eqref{lmodel} once the scalar field $\phi$ gets a VEV, as in Eq.~\eqref{VEVsDef}. Once necessary, we will move to the so-called \textit{derivative}- or \textit{chirality-preserving} basis, easily obtained by redefining the fermionic fields to rotate away the ALP from as many terms of the Yukawa sector as possible. Applying these transformations to the fermionic kinetic terms, derivative couplings between the ALP and fermion bilinears are generated. 

The common chirality-flipping Lagrangian, at LO in the $f_a$-expansion, takes the form 
\begin{equation}
    \begin{split}
        \sL_a =&\dfrac{1}{2}\partial_\mu a \partial^\mu a
        -\dfrac{1}{2}m_a^2 a^2 
        -\dfrac{ia}{f_a}\sum\limits_{ij}\left(\ov{\Psi_i}(c_{a\Psi_i\Psi_j}^L P_L+ c_{a\Psi_i\Psi_j}^R P_R)\Psi_j\right)+\\
        &-\dfrac{1}{4}g_{a\gamma\gamma}a F^{\mu\nu} \widetilde{F}_{\mu\nu}
        -\dfrac{1}{4}g_{aZZ}a Z^{\mu\nu} \widetilde{Z}_{\mu\nu}
        -\dfrac{1}{4}g_{a\gamma Z}a F^{\mu\nu} \widetilde{Z}_{\mu\nu}
        -\dfrac{1}{2}g_{aWW}a W^{+\mu\nu} \widetilde{W}_{\mu\nu}^-\,,
    \end{split}
\label{ALPLAGChiFlip}
\end{equation}
where $\Psi$ can be any of the fermionic fields of the theory.

In the following, we first focus on the ALP-mass and then we will discuss its relevant interactions for the phenomenological analysis

%%%%%%%%%%%%%%%%%%%%%%%%%%%%%%%%%
\subsubsection{The ALP mass}
\label{subsubALPMass}

As discussed in Sec.~\ref{sec:Model}, the PQ explicit breaking generates a radiative mass for the ALP after SSB. Given an interaction of a multiplet of Dirac fermions of the type
\begin{equation}
    \sL_Y\supset -\ov{\chi_L}\mathcal{M}_\chi \chi_R\,,
\end{equation}
the exact expression of the 1-loop ALP potential can be obtained through the Coleman-Weinberg (CW) potential~\cite{Coleman:1973jx}~\footnote{Notice that a factor of $1/2$ is included as Weyl fermions are employed.}, that in the $\ov{\text{MS}}$-scheme reads
\begin{equation}
\label{eq:CW-Potential}
    V_\text{CW}=-\frac{1}{32\pi^2}\left\{\Tr\left[\left(\mathcal{M}_\chi \mathcal{M}_\chi^\dagger\right)^2 \log\left(\frac{\mathcal{M}_\chi \mathcal{M}_\chi^\dagger}{\mu_R^2}\right)\right]-\dfrac32\Tr\left[\left(\mathcal{M}_\chi \mathcal{M}_\chi^\dagger\right)^2\right]\right\}\,,
\end{equation} 
with $\mathcal{M}_\chi$
is a $5\times 5$ Yukawa interaction matrix which depends, besides the Higgs, on $\phi$. Notice that only the contribution from the neutral lepton sector is relevant as the charged one does not involve any PQ breaking term; the same applies for the gauge and scalar sectors. While the exact computation of the radiative ALP mass can be found in App.~\ref{appendix:ALP-full-CW}, we report here a simplified calculation which reproduces correctly the result at LO when expanding in inverse powers of $\Lambda$ and $M_\psi$.

Through symmetry arguments discussed in Sect.~\ref{sec:Model}, we already concluded that the ALP mass must be proportional to the product of four quantities, see Eq.~\eqref{eq:ALP-mass-short}: the two Yukawa couplings $Y_V$ and $Y_{V'}$ and the two HNL mass scales, $\Lambda$ and $M_\psi$, that can be either Lagrangian masses or dynamically generated, thus proportional to $f_a$.
In particular, no LN-violating parameter is expected to contribute at LO and therefore we can set them to zero to simplify the computations. 
In this limit, we can build a Dirac pair out of the heavy neutral states defining $T\equiv N_R + S_R^c$. As a result, the relevant interactions of the Lagrangian of Eq.~\eqref{lmodel} read
\begin{equation}
\begin{split}
-\sL_Y\supset&\Lambda \overline{T_L} T_R+M_\psi\ov{\psi_L} \psi_R+Y_V \overline{T_L} \tH^{\dagger} \psi_R+Y_{V^\prime} \ov{\psi_L} \tH T_R +\hc\,, 
\end{split}
\end{equation}
corresponding to the interaction matrix
\begin{equation}
    -\sL_Y\supset \begin{pmatrix}
        \ov{T_L} & \ov{\psi_L}
    \end{pmatrix}
    \begin{pmatrix}
        \Lambda & m_V \\
        m_{V'} & M_\psi
    \end{pmatrix}
    \begin{pmatrix}
        T_R\\
        \psi_R
    \end{pmatrix}\,.
\end{equation}
 The cases in which the heavy lepton masses are generated dynamically via $\phi^{(\ast)}$ can be obtained by mapping $\Lambda\to e^{i\,\sign{x} a/f_a}\Lambda$ and $M_\psi\to e^{i\,\sign{y} a/f_a} M_\psi$, where the parameters $(x,y)$ interpolate between models A, B, C and D.
 
 The eigenvalues of such $2\times 2$ matrix can be computed easily, and the CW, in the limit of large masses of the heavy fermions, gives the ALP mass in the $\ov{\text{MS}}$-scheme
\begin{equation}
    \begin{split}
        f_a^2 m_a^2=\dfrac{(\ov{\delta}_{x,1}+\ov{\delta}_{y,1})^2}{4\pi^2}\left(\dfrac{m_V m_{V'}\Lambda M_\psi}{M_\psi^2-\Lambda^2}\right)&\left[\dfrac{(M_\psi^2+\Lambda^2)}{2}\log\left(\dfrac{M_\psi^2}{\Lambda^2}\right)\right.\\
        & \left.+(M_\psi^2-\Lambda^2)\left(\log\left(\dfrac{M_\psi\Lambda}{\mu_R^2}\right)-1\right)\right]\,,
    \end{split}
    \label{eq:massALPatLO}
\end{equation}
neglecting terms $\mathcal{O}(\Lambda^0,M_\psi^0)$, where we defined a Kronecker-delta with sign
\begin{equation}
     \ov{\delta}_{x,1}\equiv \sign{x}\delta_{|x|,1}\,,\qquad\qquad
     \ov{\delta}_{y,1}\equiv \sign{y}\delta_{|y|,1}\,.
\end{equation}
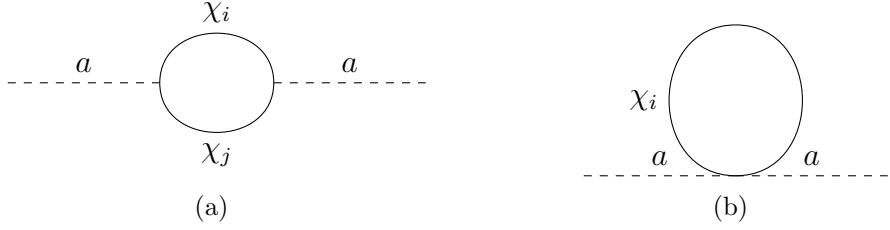
\begin{figure}[]
\centering
\subfigure[{}\label{fig:bubble}]{
\begin{tikzpicture}
  \begin{feynman}
    \vertex (a) ;
    \vertex at ($(a) + (2cm, 0cm)$)(b) ;
    \vertex at ($(b) + (1.5cm,0cm)$)(c) ;
    \vertex at ($(c) + (2cm, 0cm)$)(d) ;    
    \diagram* {
      (a) -- [scalar,edge label=\(a\)] (b);
      (c) -- [scalar,edge label=\(a\)] (d);
      (b) --[plain, edge label=\(\chi_i\), looseness=1.2, half left] (c) --[plain, edge label=\(\chi_j\), half left] (b);
    };
  \end{feynman}
\end{tikzpicture}
}
\qquad \qquad 
\subfigure[{}\label{fig:balloon}]{
\begin{tikzpicture}
  \begin{feynman}
     \vertex (z) ;
    \vertex at ($(z) + (0cm, 3cm)$)(a) ;
    \vertex at ($(a) + (2cm, 0cm)$)(b) ;
    \vertex at ($(b) + (2cm, 0cm)$)(c) ;  
    \vertex at ($(a) + (2cm, 2cm)$)(d) ;
    \diagram* {
      (a) -- [scalar,edge label=\(a\)] (b);
      (b) -- [scalar,edge label=\(a\)] (c);
      (b) --[plain, edge label=\(\chi_i\), half left] (d) -- [plain, half left] (b);
    };
  \end{feynman}
\end{tikzpicture}
}
\caption{Representative set of Feynman diagrams contributing to the ALP mass at 1-loop where $\chi$ stands for the neutral lepton multiplet, defined in Eq.~\eqref{multiplet}.}
\label{fig:ma-diagrams}
\end{figure}
In general, the diagrammatic computation of the ALP mass involves the two types of diagrams shown in Fig.~\ref{fig:ma-diagrams}. As detailed in Ref.~\cite{deGiorgi:2023tvn}, the LO PQ-breaking contribution to the mass can be conveniently calculated in the derivative basis, computing only the second diagram.
Performing the explicit computation of the different 1-loop diagrams with a single heavy lepton exchange, the ALP mass can be decomposed as the sum of two contributions~\cite{deGiorgi:2023tvn}
\be
f_a^2\, m_a^2= f_a^2\left(m_{a,\psi}^2+m_{a,N/S}^2\right)\,,
\ee
where
\be
\begin{aligned}
    f_a^2\,m_{a,\psi}^2&=\dfrac{(\ov{\delta}_{x,1}+\ov{\delta}_{y,1})^2}{4\pi^2}\left(\dfrac{m_V m_{V'}\Lambda M_\psi}{M_\psi^2-\Lambda^2}\right)\times M_\psi^2\left[\log\left(\dfrac{M_\psi^2}{\mu_R^2}\right)-1\right]\,,\\
    f_a^2\,m_{a,N/S}^2&=-\dfrac{(\ov{\delta}_{x,1}+\ov{\delta}_{y,1})^2}{4\pi^2}\left(\dfrac{m_V m_{V'}\Lambda M_\psi}{M_\psi^2-\Lambda^2}\right)\times\Lambda^2\left[\log\left(\dfrac{\Lambda^2}{\mu_R^2}\right)-1\right]\,,
\end{aligned}
\ee
and gives the same result as the CW approach. 

This LO expression for the ALP mass vanishes for $M_\psi=\Lambda$ and the NLO contributions should be considered. In this specific limit, the value of the ALP mass reads
\be
f_a^2\,m_a^2=(\ov{\delta}_{x,1}+\ov{\delta}_{y,1})^2\frac{m_V m_{V^\prime} \left(2 m_{V^\prime}^2+3m_N^2-2m_{V^\prime}m_V+2 m_V^2\right)}{24\pi^2}
\label{eq:ALP-mass-NLO}
\ee
where only terms of order $\order{\Lambda^{-1}}$ have been considered.
The complete expression will be considered in the numerical analysis.

It is interesting to notice that the Yukawa couplings cannot take arbitrary signs, but the ALP potential dynamically fixes the product $Y_V\times Y_{V'}>0$ (analogue to Ref.\cite{deGiorgi:2023tvn}) to ensure a positive ALP mass. 

\begin{figure}[h!]
	\centering
\subfigure[{}\label{fig:renorm-scale-dep}]{\includegraphics[width=0.48\linewidth]{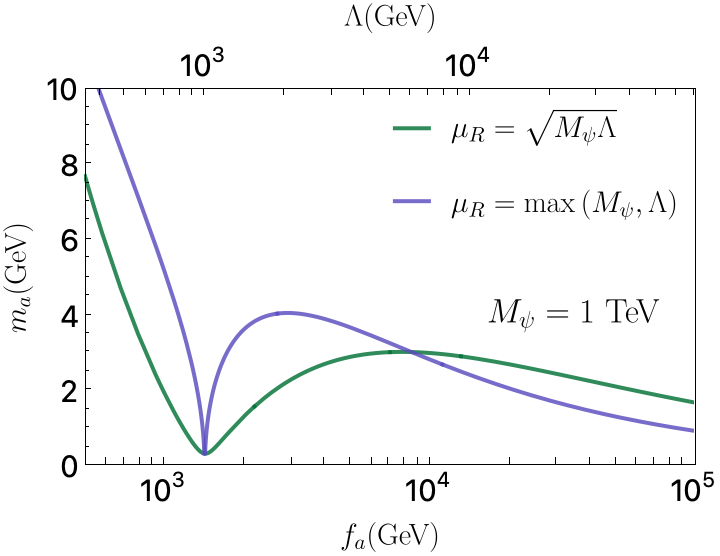}}
     \,
     \subfigure[{}\label{fig:mALP-modelA}]{\includegraphics[width=0.48\linewidth]{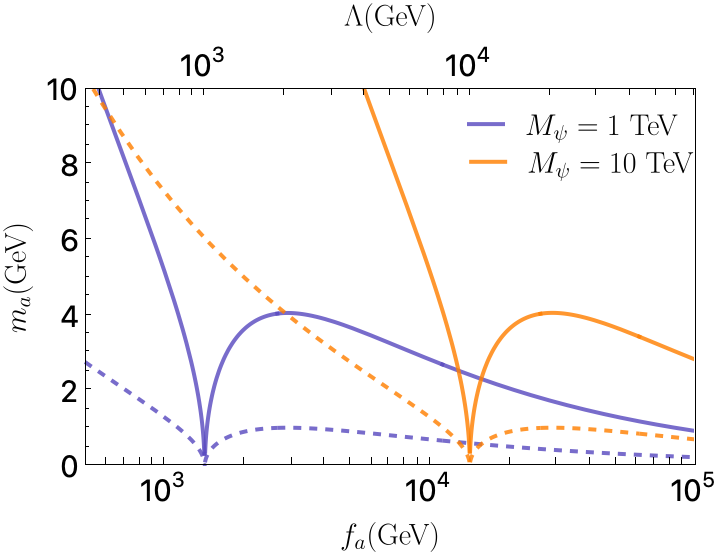}}
     \caption[]{\em ALP mass in $\overline{\text{MS}}$-scheme as a function of $\Lambda$ and $f_a$ for model A. $\Lambda$ and $f_a$ are related via $\Lambda=\alpha_N f_a/\sqrt{2}$. For convenience, $\alpha_N=1$. As a reference value, $Y_N=0.1$. \textbf{Left:} Dependence on the renormalisation scale of the ALP mass. $Y_V=Y_{V^\prime}=0.4$. The green line indicates $\mu_R=\sqrt{M_\psi \Lambda}$, while the blue represents $\mu_R=\max(M_\psi,\,\Lambda)$. The minimum corresponds to the point where $\Lambda=M_\psi$. \textbf{Right:} ALP mass as a function of $f_a$ and $\Lambda$ for different values of $M_\psi$. $Y_V=Y_{V^\prime}=0.4~(0.1)$ for the continuous~(dashed) line.}
	\label{fig:mALPfa}
\end{figure}

\begin{figure}[h!]
   \centering
    \subfigure[{}\label{fig:ratioalpha1}]{\includegraphics[width=0.48\linewidth]{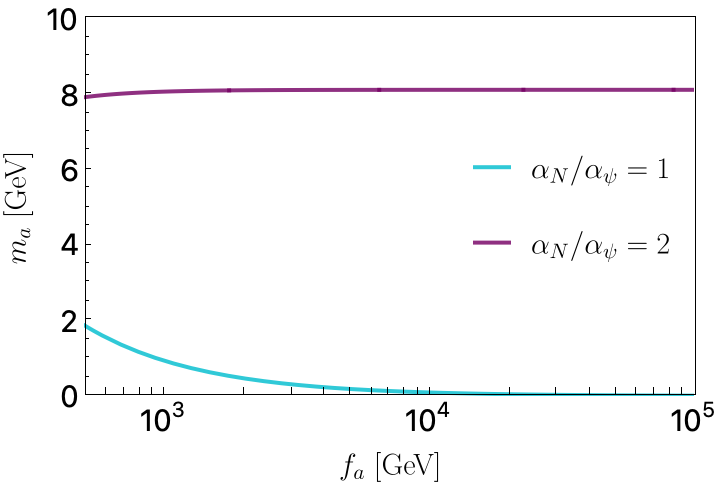}}
     \,
     \subfigure[{}\label{fig:ratioalpha2}]{\includegraphics[width=0.48\linewidth]{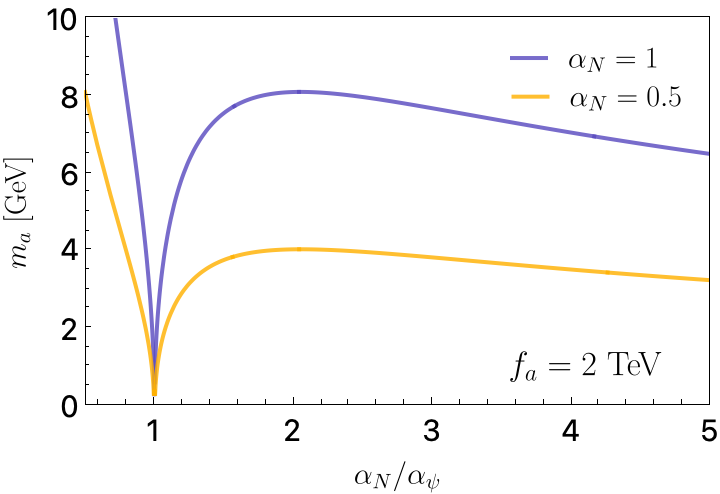}}
    \caption{\em ALP mass at NLO (see Eq.~\eqref{CWgeneral}) in $\overline{\text{MS}}$-scheme for model C. In both cases, $Y_V=Y_{V\prime}=0.4$. \textbf{Left:} Dependence on the scale $f_a$ for two different values of the ratio $\alpha_N/\alpha_\psi$. The cyan (purple) line corresponds to the case in which $\alpha_N/\alpha_\psi=1\left(2\right)$. In both cases, $\alpha_N=1$. $Y_N=0.1$ is taken as a reference value. \textbf{Right:} ALP mass as a function of the ratio $\alpha_N/\alpha_\psi$ for two values of $\alpha_N=0.5, 1$, which represent the orange and blue lines, respectively. $f_a$ has been fixed to $2 \TeV$.}
    \label{fig:ratioalpha}
\end{figure}

The dependence of $m_a$ on $\Lambda$ and $M_\psi$ can be seen in Fig.~\ref{fig:mALPfa} for model A regardless of the presence of $\phi$ or $\phi^\ast$ in the Lagrangian: indeed $m_a^2\propto (\ov{\delta}_{x,1})^2$. The same behaviour is also obtained for model B at LO as the roles of $M_\psi$ and $\Lambda$ can be interchanged in Eq.~\eqref{eq:massALPatLO}, that is $m_a^2\propto (\ov{\delta}_{y,1})^2$. On the other hand, the mass dependence on $f_a$ for model C is different and is discussed in Fig.~\ref{fig:ratioalpha}.

Fig.~\ref{fig:renorm-scale-dep} shows the dependence of the ALP mass on the renormalisation scale, very much typical of the situation where more than one heavy scale is present: the green line corresponds to $\mu_R=\sqrt{M_\psi\Lambda}$ that estimates an average mass scale between $M_\psi$ and $\Lambda$; in the blue line instead $\mu_R$ is taken to be the largest scale among $M_\psi$ and $\Lambda$. Both lines have a local minimum corresponding to $M_\psi=\Lambda$, where the LO contribution vanishes. The discontinuity of the derivative of the blue line is simply due to the change of value for $\mu_R$: up to $\Lambda=1\TeV$ we have $\mu_R=M_\psi$, while above that value $\mu_R=\Lambda$. In Fig.~\ref{fig:mALP-modelA}, we fix $\mu_R=\max(M_\psi,\Lambda)$, but vary the value of $M_\psi$ and of $Y_V=Y_{V'}$ to show the dependence of the ALP mass on these parameters. Increasing the value of $M_\psi$ we obtain a larger ALP mass, but at the price of a hierarchy between $\Lambda$ and $M_\psi$. On the other hand, lowering the value of the two Yukawa couplings from $Y_V=Y_{V'}=0.4$ to $0.1$ suppresses the result of the ALP mass, as we expect from the expression analysis. In both plots, $\alpha_N$ is taken to be equal to 1 for simplicity, and thus $\Lambda$ is smaller than $f_a$ just by a $\sqrt2$ factor: a stronger hierarchy between these two scales is achievable for $\alpha_N<1$, although values $\alpha_N\ll1$ are not expected from a symmetry invariance point of view. The value of $Y_N$ is fixed at $0.1$, although larger values are viable for larger $f_a$. As we will discuss in the next section, these values are consistent with other experimental bounds. 

Fig.~\ref{fig:ratioalpha} illustrates the prediction for $m_a$ in model C, where the dependence is different as both $M_\psi$ and $\Lambda$ are obtained via the SSB of the $U(1)_{PQ}$. In particular, for $\alpha_N/\alpha_\psi=1$ the two masses coincide, $M_\psi=\Lambda$, and thus the LO contribution to the ALP mass vanishes and the expression in Eq.~\eqref{eq:ALP-mass-NLO} holds. This case is shown in Fig.~\ref{fig:ratioalpha1}, where the cyan line illustrates the $m_a$ dependence on $f_a$. On the other hand, $\alpha_N/\alpha_\psi\neq1$, the dependence on $f_a$ completely disappears:
\begin{equation}
    \begin{split}
        m_a^2=
        \dfrac{\alpha_\psi^2}{\pi^2}
        \left(
        \dfrac{m_V m_{V'}\frac{\alpha_N}{\alpha_\psi}}{1-\frac{\alpha^2_N}{\alpha^2_\psi}}
        \right)
        &\left[
        \left(
        1+
        \dfrac{\alpha_N^2}{\alpha_\psi^2}\right) \log\left(\dfrac{\alpha_\psi}{\alpha_N}\right)+
        \left(1-\dfrac{\alpha_N^2}{\alpha_\psi^2}\right)\left(\log\left(\dfrac{\alpha_\psi\alpha_N}{\max(\alpha_\psi,\alpha_N)}\right)
        -1
        \right)
        \right]\,,
    \end{split}
    \label{eq:massALPatLO-alpha}
\end{equation}
taking as renormalisation scale $\mu_R=\max(\alpha_\psi,\,\alpha_N)f_a/\sqrt2$.
The independence of $m_a$ from $f_a$ can be seen in Fig.~\ref{fig:ratioalpha1}, where the horizontal purple line corresponds to $\alpha_N/\alpha_\psi=2$. The small deviation for small $f_a$ values is due to the NLO contribution to $m_a$ (see Eq.~\eqref{CWgeneral}). The Yukawa $Y_N$ is fixed at $0.1$, consistently with the previous plots in Fig.~\ref{fig:mALPfa}.
Finally, the dependence of $m_a$ on the ratio $\alpha_N/\alpha_\psi$ is shown in Fig.~\ref{fig:ratioalpha2}, for two different values of $\alpha_N=0.5,\,1$.\\ 

Higher-order loop corrections with dynamical Higgs can result in enhancements of order $(\Lambda,M_\psi)^2/v^2$ in the same fashion as the Higgs hierarchy problem.
Even if one takes into account such corrections, in our models the largest scale is of order TeV, and therefore at worst the mass contribution from 2-loops is expected to be

\begin{equation}
\label{eq:2-loops}
    \delta m_{a,\text{2-loops}}^2\sim m_\text{a,1-loop}^2\times \dfrac{1}{16\pi^2}\left(\dfrac{1~\text{TeV}}{v}\right)^2\,\log\left(\dfrac{1\,\text{TeV}^2}{M_W^2}\right) \approx 0.5 \times m_{a,\text{1-loop}}^2\,.
\end{equation}

The 2-loops contribution is expected to be a $\mathcal{O}(1)$ correction to the result we obtained, thus explaining the relatively large dependence of our result on the choice of $\mu_R$~(cfr.~Fig.~\ref{fig:mALPfa}). The three loops contribution is expected to be much smaller as no overall rescaling as in Eq.~\eqref{eq:2-loops} is possible; further loops are expected to contribute with
\begin{equation}
    \dfrac{1}{16\pi^2}\,\log\left(\dfrac{1\,\text{TeV}^2}{v^2}\right) \sim 2\%\,,
\end{equation}
thus not endangering the perturbative loop expansion beyond the two-loop result. Therefore, we do not expect these corrections to change qualitatively, but only quantitatively, any of the results.
A precise estimation of the ALP mass requires careful analysis of the two-loops contributions. However, as the model only deals with the second generation of leptons, an extension to three families is necessary to complete the description: it is expected to possibly quantitatively change the result, jeopardising the effort to compute the two-loop correction in our simplified context. We therefore do not delve further into such matter and use the 1-loop result as a benchmark, in order to provide a qualitative description of the output of the model.\\

All in all, the ALP in these models receives contributions of $\mathcal{O}(\text{GeV})$, which allows it to escape the stringent astrophysical constraint.

%
%%%%%%%%%%%%%%%%%%%%%%%%%%%%%%%%%%%%%%%%%%%%%%%%%%%%%%%%%%%
\subsubsection{ALP interactions}
Let us focus now on the interactions of the ALP. A comprehensive list of all tree-level interactions in the mass basis at LO can be found in App.~\ref{appendix:Lagrangia-mass-basis}. Besides the coupling to gauge bosons, the decay $a\to \mu^+\mu^-$ is kinematically viable since the ALP-mass is of $\mathcal{O}(\text{GeV})$ and a coupling with muons, $a\ov{\mu_L}\mu_R$, could generate strong bounds on the parameter space of the model. We consider them separately in the following.

\paragraph{ALP-Gauge Bosons Interactions:}
The couplings with gauge bosons are generated at 1-loop via the anomalous triangle diagram of the $\psi^-$. A non-vanishing result requires chiral PQ-charge assignments for $\psi^-_L$ and $\psi^-_R$, which is always achieved in our models if $M_\psi$ is generated via SSB, that is in models B, C, and D. By explicit computation, the couplings to on-shell gauge bosons are found to be
\begin{align}
    &g_{a\gamma\gamma}=\ov{\delta}_{y,1}\frac{\alpha_\text{em}}{\pi f_a}\,,&&g_{aZZ}=\ov{\delta}_{y,1}\frac{\alpha_\text{em}}{6\pi f_a s_{2\theta_W}^2}\left(c_{4\theta_W}+7\right)\,,&&g_{aWW}=\ov{\delta}_{y,1}\frac{\alpha_\text{em}}{2\pi f_a s_{\theta_W}^2}\,,
    \label{eq:vector-coupling}
\end{align}
where $\theta_W$ is the Weinberg angle and $s_X[c_X]\equiv \sin \left[\cos \right](X)$. The coupling $g_{aZ\gamma}$ is found to be zero at LO in the expansion.

\paragraph{ALP-Muons Interactions:}
As there are no direct terms in the interaction basis that couple $\phi$ to SM fields, they can only be generated by mixing when rotating to the mass basis. However, such contributions do not arise in any of the considered models. This can be first understood by looking at the diagonalisation of the charged mass matrix: as only a RH fields' rotation is required~(cfr.~Eq.~\eqref{eq:charged-mixing}), and as the ALP couples only to $\psi^-$ in the interaction basis, there is no possibility to generate a coupling of the form $a\ov{\widehat{\mu}_L}\widehat{\mu}_R$ at any order at tree-level, leaving at best
\begin{equation}
\label{eq:ALP-psi-mu-int}
    \sL_a\supset i\frac{a}{f_a}\,\left(\ov{\delta}_{y,1}m_R \ov{\widehat{\psi}^-_L}\widehat{\mu}_R\right)+\hc\,.
\end{equation}

\begin{figure}[tbh]
	\centering
    \includegraphics[width=0.9\linewidth]{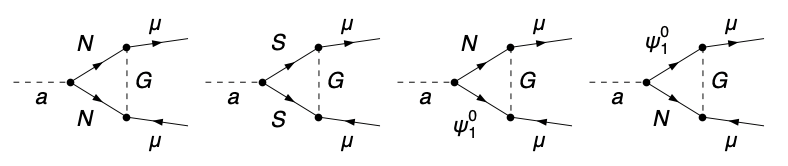}%}
	\caption[]{\em Illustrative set of triangle diagrams that contribute at 1-loop to $a\to \mu^+\mu^-$. $N,\,S,\,\psi_1^0$ are Majorana fermions defined in App.~\ref{appendix:Lagrangia-mass-basis}.}
	\label{fig:triangles}
\end{figure}

The situation changes when considering the loop-level contributions. There are two possible contributions to the coupling $a\ov{\widehat{\mu}_L}\widehat{\mu}_R$, namely indirectly induced via 1-loop mixing of $\mu-\psi$~(cfr.~Eq.~\eqref{eq:self-energy-mix}) from the interaction of Eq.~\eqref{eq:ALP-psi-mu-int}, or direct via triangle diagrams, as exemplified in Fig.~\ref{fig:triangles}.

The total contribution stemming from the triangle diagrams, $iV_{a\mu\mu}$, in the  $\ov{\text{MS}}$-scheme reads
\begin{align}
    iV_{a\mu\mu}&(P_R-P_L)= i\frac{ m_N m_R m_V}{8 \pi ^2 f_a v^2}\left(\frac{\Lambda}{M_\psi}\right)\times\frac{P_R-P_L}{ \left(M_\psi^2-\Lambda ^2\right)}\times \\
    &\nonumber  \times\left[M_\psi^2 (\ov{\delta}_{x,1}+\ov{\delta}_{y,1}) \log \left(\frac{\mu_\text{R} ^2}{M_\psi^2}\right)-\log \left(\frac{\mu_\text{R} ^2}{\Lambda ^2}\right) \left(\ov{\delta}_{x,1} \Lambda ^2+\ov{\delta}_{y,1} M_\psi^2\right)+\ov{\delta}_{x,1} \left(M_\psi^2-\Lambda ^2\right)\right] \,,
\end{align}
where $P_{R,L}=(1\pm\gamma_5)/2$ are the right(left)-chirality projectors. Including the mixing between $\mu$ and $\psi^-$ from Eq.~\eqref{eq:left-mixing}, the total ALP-muons coupling as defined in Eq.~\eqref{eq:ALP-psi-mu-int} is given by
\begin{align}
\label{eq:loop-ALP-mu}
        c_{a\mu\mu}^R&=-c_{a\mu\mu}^L = \left(m_R\frac{\delta M_L}{M_\psi}+V_{a\mu\mu}\right)\,,\\
        &\nn= \left(\ov{\delta}_{x,1}+\ov{\delta}_{y,1}\right)\dfrac{m_N\,m_R\,m_V}{8\,\pi^2\,v^2}
    \left(\dfrac{\Lambda}{M_\psi}\right)
    \left[1+\dfrac{1}{M_\psi^2-\Lambda^2}\left(M^2_\psi
    \log{\dfrac{\mu_\text{R}^2}{M_\psi^2}}-
    \Lambda^2
    \log{\dfrac{\mu_\text{R}^2}{\Lambda^2}}\right)\right]\,,\\
    &\nn = \left(\ov{\delta}_{x,1}+\ov{\delta}_{y,1}\right)\times \delta m_\mu\Big|_{m_{V'}=0}\,,
\end{align}
where with $\delta m_\mu\Big|_{m_{V'}=0}$ we refer to the muon mass obtained in Eq.~\eqref{deltammu} and computed for $m_{V'}=0$. In an exact PQ-symmetric model, the ALP-muon coupling would be proportional to the pole-mass of the muon; in our case, this is not exactly the case as the symmetry is explicitly broken by the simultaneous presence of $Y_{V}$ and $Y_{V'}$. We further investigate this aspect in App.~\ref{ALPMuonsCouplings}.

%%%%%%%%%%%%%%%%%%%%%%%%%%%%%%%%%%%%%%%%%%%%%%%%%%%%%%%%%%%
%%%%%%%%%%%%%%%%%%%%%%%%%%%%%%%%%%%%%%%%%%%%%%%%%%%%%%%%%%%
%
\section{Phenomenological analysis}
\label{sec:Pheno}
\begin{figure}[tbh]
    \centering
    \includegraphics[width=0.8\textwidth]{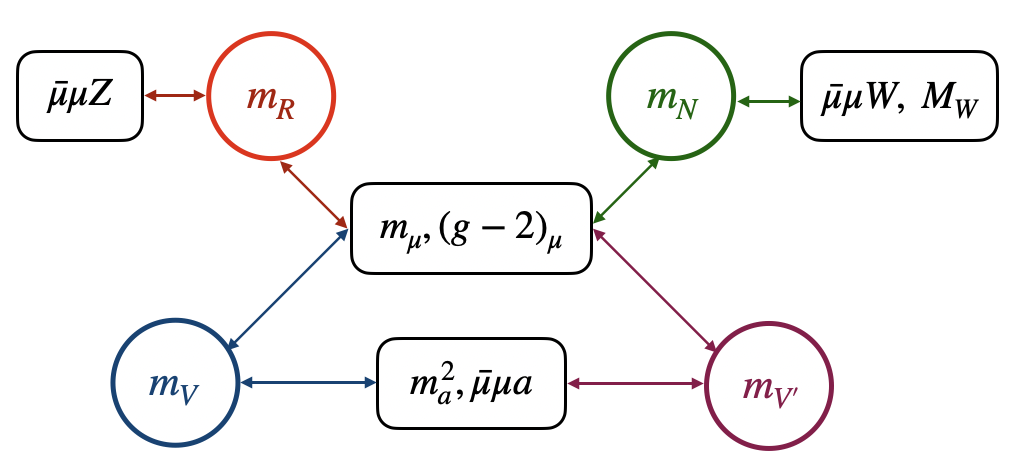}
    \caption{\em Couplings and connections to the relevant observables.}
    \label{fig:scheme}
\end{figure}

In this section, we examine the relevant phenomenology to constrain the couplings and predict the masses of the heavy fermions and the ALP. In Fig.~\ref{fig:scheme} we report the scheme of the connection, up to 1-loop, between the couplings and the observables considered in this section. The bounds limited to the tree-level SMEFT analysis can be found in Ref.~\cite{Crivellin:2021bkd}.

The input parameters used for the numerical analysis are~\cite{Workman:2022ynf}:
\be
\begin{aligned}
m^\text{exp}_\mu&=105.6583755(23)\MeV \\
\alpha_\text{em}&=7.2973525693(11)\times10^{-3}\\
G_\mu&=1.1663788(6)\times10^{-5}\GeV^{-2}\\
M_Z&=91.1876(21)\GeV
\end{aligned}
\ee
We will not discuss the phenomenology of model D in what follows, as we are only interested in the models where the ALP is massive.

%
%%%%%%%%%%%%%%%%%%%%%%%%%%%%%%%%%%%%%%%%%%%%%%%%%%%%%%%%%%%
%
\subsection{Relevant bounds}

\paragraph{\boldmath$m_N/\Lambda$:}
This combination of parameters is constrained by a series of observables. The first one is the determination of the $W$-gauge boson mass. The most recent and precise value of $M_W$ was obtained by the ATLAS collaboration~\cite{ATLAS:2023fsi}, $M_W=80.360(16)\GeV$~\footnote{In April 2022 the CDF collaboration reported the most precise measurement value of $M_W$~\cite{CDF:2022hxs}, $M_W^\text{CDF II}=80.433(9)\GeV$. Such a measurement is in tension with previous determinations with significance of $7.0\sigma$ and is nowadays subject of controversy within the experimental community to establish its accuracy. Its impact in the analysis of the HNLs contributions has been studied in Ref.~\cite{deGiorgi:2022xhr}. In the following, waiting for the matter to be settled, we will adopt a more conservative approach and use the most precise value $M_W$ compatible with the world average pre-CDF~II measurement.}. 
The modification of the coupling of the $W$-boson with muons induces a modification in the muon decay $\mu\to e\ov{\nu}\nu$. The measurement of the muon lifetime allows for the most precise measurement of the Fermi constant. The modified decay width of the muon reads
\begin{equation}
    \Gamma_\mu \simeq \dfrac{(m_\mu^\text{exp})^5 G_F^2}{192\,\pi^3}\left(1-\dfrac{\widetilde m_N^2}{2\Lambda^2}\right)^2\equiv\dfrac{(m_\mu^\text{exp})^5 G_\mu^2}{192\,\pi^3}\,,
\end{equation}
where $G_F$ is the Fermi constant as defined in the Fermi Lagrangian and $G_\mu$ is its experimental determination extracted from the muon lifetime such that
\begin{equation}
    G_F 
    \simeq G_\mu \left(1+\dfrac{\widetilde m_N^2}{2\Lambda^2}\right) \,.
\end{equation}
As the most precise determination of the $M_W$ takes as input parameters $\alpha_\text{em}$, $M_Z$ and $G_\mu$, the prediction for the $M_W$ gets shifted
\begin{equation}
    M_W\simeq M_Z\,
    \sqrt{
    \dfrac{1}{2}+
    \sqrt{
    \dfrac{1}{4}-
    \dfrac{\pi\,\alpha_\text{em}}{\sqrt2\,G_\mu\,M_Z^2\,(1-\Delta r)}
    \left(1-\dfrac{\widetilde m_N^2}{2\Lambda^2}\right)
    }}
    \,,
    \label{MWPrediction1Gen}
\end{equation}
where $\Delta r=0.03657(21)(7)$ takes into account radiative electromagnetic corrections and the on-shell scheme has been used, thus fixing the Weinberg angle to be
\be
\sin\theta_W\equiv1-\dfrac{M_W^2}{M_Z^2}\,.
\ee

Considering the latest ATLAS measurement\cite{ATLAS:2023fsi} of $M_W$ we thus obtain the following upper bound on $m_N$
\be
\left(\dfrac{m_N}{\Lambda}\right)^2\leq 4.0 \cross 10^{-3}\,.
\label{eq:ratioN}
\ee

This same combination can be constrained by the invisible $Z$-decay rate and from Lepton Flavour Universality violating processes. However, as discussed in Ref.~\cite{deGiorgi:2022xhr}, once extending this setup to more realistic models with three generations, these constraints get relaxed and the most relevant remains the one in Eq.~\eqref{eq:ratioN}.

Other bounds on this combination arise from direct searches of HNLs at colliders. As shown in Ref.~\cite{deGiorgi:2022xhr}, the present corresponding bounds are two orders of magnitude weaker than the one from the $M_W$ determination. 

One might wonder whether loop corrections to the $W$ mass involving loops of HNLs or ALP couplings are relevant. Both corrections to the self-energy of the $W$ (or $Z$) are expected to be $\Pi_{W,Z}\propto v^2/(16\pi^2 \Lambda^2)$ (or alternatively $M_\psi$ or $f_a$, instead of $\Lambda$). All in all, these contributions are expected to be subleading compared to the tree-level modification of Eq.~\eqref{MWPrediction1Gen} by a loop factor, as stated in Ref.~\cite{deGiorgi:2022xhr}.
%
%%%%%%%%
\paragraph{\boldmath $m_R/M_\psi$:}
The coupling $Y_R$ modifies the coupling strength of the Z-boson to right-handed muons. Its deviation can be constrained by EW-fit data and at $2\sigma$ reads~\cite{deGiorgi:2022xhr}
\begin{equation}
	\left(\dfrac{m_R}{M_\psi}\right)^2 \leq 5.6 \cdot 10^{-3}\,.
	\label{eq:ratioR}
\end{equation}

As discussed in Ref.~\cite{deGiorgi:2022xhr}, this bound is much stronger than the one we can extract on the same quantity from collider searches on both Higgs couplings and direct searches on HNLs, which are two orders of magnitude weaker than Eq.~\eqref{eq:ratioR}.

\paragraph{\boldmath $\epsilon Y_S$ from neutrino masses:}
As this simplified model deals exclusively with one leptonic generation, only one active neutrino will acquire a mass via the Seesaw mechanism. Choosing to reproduce the atmospheric mass splitting, $\Delta m_{\text{atm}}^2\sim2.5\times 10^{-3}\eV^2$~\cite{Esteban:2020cvm}, we obtain
\ba
&\widehat{m}_\nu=-\frac{2 \epsilon m_N m_S}{\Lambda}\approx \sqrt{\abs{\Delta m_{\text{atm}}^2}}\,,&&\Rightarrow& %(\epsilon Y_S^2 \gtrsim 1.6\cdot10^{-25} 
\abs{\frac{\epsilon Y_S Y_N}{\Lambda}}\sim 8.3\cdot 10^{-13} \TeV^{-1}\,. 
\ea

%%%%%%%%%%%%%%%%%%%%%%%%%%%%%%%%%%%%%%%%%%%%%%%%%%%
%
\boldmath
\subsection{Contributions to the $(g-2)_\mu$}
\unboldmath

In the model we consider, both the muon mass and anomalous magnetic moment are generated at 1-loop. Once focusing on the HNL contributions, they must receive contributions that are proportional to the same couplings, as the only difference in their Feynman diagrams relies on including or not the external photon leg,
\begin{equation}
    \delta m_\mu \propto \delta a_\mu \propto Y_N Y_R Y_{V,V'}\,. 
\end{equation}

The latest and most precise measurement of the $(g-2)_\mu$ comes from the Muon g-2 Collaboration~\cite{Muong-2:2023cdq}
\begin{equation}
    a_\mu\equiv \dfrac{g_\mu-2}{2}=116592059(22)\cross 10^{-11}\,,
\end{equation}
yielding, according to the Muon $g-2$ Theory Initiative~\cite{Aoyama:2020ynm}, a discrepancy with respect to the SM prediction of
\begin{equation}
     \delta a_\mu^\text{exp}\equiv a_\mu^\text{exp}-a_\mu^\text{SM}= (2.49\pm 0.49)\cross 10^{-9}\,,
     \label{eq:g-2-discrepancy}
\end{equation}
corresponding to a significance of $5.1\sigma$. Some comments regarding the above result are in order as it is currently subject to several controversies. The main source of uncertainty to the SM prediction, $a_\mu^\text{SM}$, comes from non-perturbative contributions of hadronic-vacuum polarization~(HVP). The Muon $g-2$ Theory Initiative~\cite{Aoyama:2020ynm} prediction has been performed by extracting such contribution from low-energy $e^+e^-\to \text{hadrons}$ cross sections employing the so-called \textit{dispersive} method.
In April 2021 the BMW collaboration first reported a lattice result with the first sub-percent error in the estimation of the HVP~\cite{Borsanyi:2020mff}, making lattice results competitive with dispersive uncertainties for the first time. Such a result is in tension by $\sim 2\sigma$ with dispersive results and relaxes the $\delta a_\mu^\text{exp}$ significance to $1.7\sigma$. Other collaborations afterwards independently managed to achieve similar precision in computing partial contributions to HVP, being able to partially counter-check the BMW result. While the so-called \textit{short-distance} lattice contribution to HVP seems to agree with the dispersive method~\cite{Colangelo:2022vok} (see e.g. Ref.~\cite{Kuberski:2024bcj}), the so-called \textit{intermediate-distance} lattice contribution~\cite{ExtendedTwistedMass:2022jpw, Ce:2022kxy, RBC:2023pvn} was found to be larger than the result based on the dispersive method with $\sim 4\sigma$ tension, thus supporting the full BMW result. Finally, in February 2023 the CDM-3 collaboration released a new measurement of the cross-section $e^+ e^-\to \pi^+\pi^-$~\cite{CMD-3:2023alj}, obtaining a dispersive partial contribution to HVP significantly larger than all previous estimations with significance $\gtrsim 3\sigma$, thus decreasing $\delta a_\mu^\text{exp}$ discrepancy to $2.4\sigma$. However, at present, this measurement remains a stand-alone and therefore requires further investigations.

The situation is therefore extremely controversial and far from being settled. We nevertheless take the discrepancy of Eq.~\eqref{eq:g-2-discrepancy} as input for the analysis. If in the future the solution to such a dispute was found to bring the SM prediction in agreement with the experimental result, the $(g-2)_\mu$ would serve as one of the strongest bounds on the model here studied.

In the SM, all contributions to $a_\mu$ are suppressed by two powers of $m_\mu^\text{exp}$. The first one is unavoidable and comes from the definition of the magnetic dipole operator,
\begin{equation}
    \dfrac{1}{2m_\mu^\text{exp}}\ov{\mu_L}\sigma^{\mu\nu}\mu_R F_{\mu\nu}\,.
\end{equation}
The second is peculiar to the SM and is due to the chiral structure of the operator and hence is typically dubbed as ``chirally suppressed''~(CS). In fact, the SM possesses approximate chiral symmetry per lepton generation broken explicitly solely by the Yukawa interactions which guarantees the presence of the muon mass suppression to $a_\mu$ at each order in perturbation theory~\footnote{The chiral symmetry receives extra breaking sources in the presence of non-vanishing neutrino masses, which are nevertheless completely negligible at this level of precision.}. 

Generic BSM constructions typically face the same suppression and therefore, once attempting to solve the mentioned tension, the scale of NP is relatively low, often severely constrained by other experiments. The same is not true in models where the new fields introduce extra breaking of the chiral symmetry and are hence dubbed ``chirally enhanced''~(CE). In practice, one needs a series of interactions that bridge between LH and RH components of the muon field. Such models have been systematically classified in Refs.~\cite{Kannike:2011ng,Dermisek:2013gta,Arcadi:2021cwg,Lu:2021vcp,Guedes:2022cfy} at LO. 

In our model, such CE contribution to $a_\mu$ is present only at 1-loop and mediated by HNLs, as the muon mass is also generated at 1-loop. Assuming $\Lambda,\,M_\psi \gg v$, based on power counting, one would naively expect the LO contribution to be
\begin{equation}
    \delta a_\mu^\text{naive}\propto \dfrac{Y_N Y_R Y_{V,V'}}{16\pi^2}\dfrac{m_\mu^\text{exp}v}{\Lambda M_\psi}+\mathcal{O}\left(\dfrac{1}{(M_\psi\Lambda)^2}\right)\,,
\end{equation}
thus requiring the NP scale to be at $\mathcal{O}(10)\TeV$. Remarkably, an accidental cancellation unique to this model cancels this contribution, and the first non-vanishing term reads
\be
  \delta a_\mu=
    \dfrac{3\,m^\text{exp}_\mu}{4\,\pi^2\,v^2}
    \dfrac{M_W^2}{\Lambda M_\psi}
    \dfrac{m_N m_R}{{M}_\psi}
    \left(\dfrac{m_V}{M_\psi}+\dfrac{m_{V'}}{\Lambda}\right)
    F_0\left(\dfrac{\Lambda^2}{M_W^2},\dfrac{M_\psi^2}{M_W^2}\right)\,,
\label{g2muCE1LNLO}    
\ee
 where the loop function is defined by
\begin{equation}
    F_0(x,y)\equiv \dfrac{3}{2}-\dfrac{x\log{y}-y\log{x}}{x-y}\,.
\end{equation}
This allows to lower the NP scale down to $\mathcal{O}(1)\TeV$.
Such a feature was first pointed out for its phenomenological significance in Ref.~\cite{Arkani-Hamed:2021xlp} and further studied in detail in Refs.~\cite{Craig:2021ksw,DelleRose:2022ygn, deGiorgi:2022xhr}. 

 \begin{figure}[h!]
	\centering
		\includegraphics[width=0.5\linewidth]{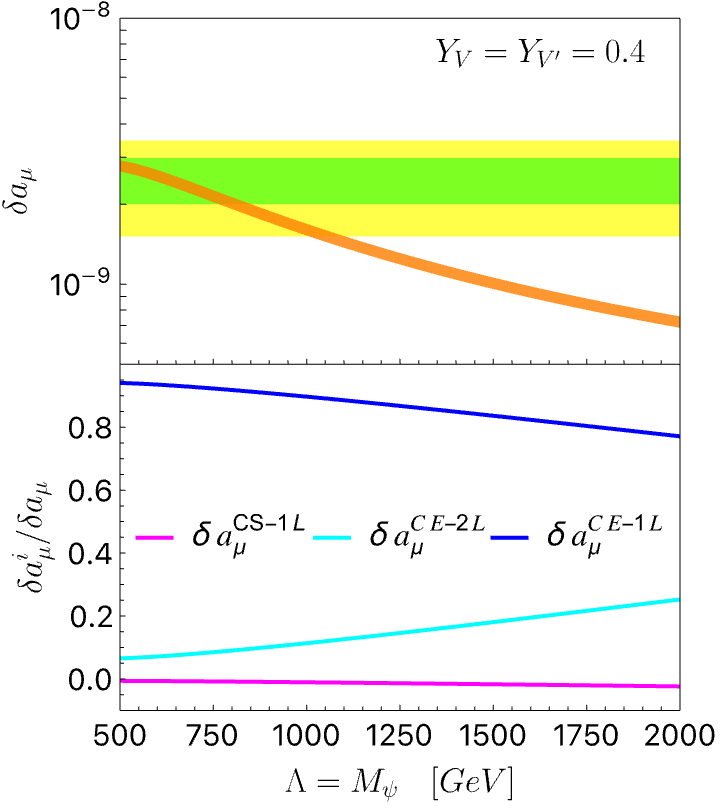}
	\caption[]{\em 
 Illustrative example of $\delta a_\mu$ as a function of $\Lambda=M_\psi$ for $Y_V=Y_{V^\prime}=0.4$. $m_N/\Lambda$ and $m_R/M_\psi$ are taken to saturate the bounds of Eqs.~\eqref{eq:ratioN}, \eqref{eq:ratioR}, respectively. \textbf{Above:} $\delta a_\mu$ (orange) and the experimental value at $1\sigma$ ($2\sigma$) in green (yellow). \textbf{Below:} ratios of the different components $\delta a_\mu^{\text{CS-1L}}$ (magenta), $\delta a_\mu^{\text{CE-2L}}$ (cyan) and $\delta a_\mu^{\text{CE-1L}}$ (blue) to the total contribution $\delta a_\mu$.}
	\label{fig:equalATLAS}
\end{figure}

An illustrative example of the possible NP contribution to $a_\mu$ is shown in Fig.~\ref{fig:equalATLAS}, where we also report the relative contributions of the 2-loops CE-contributions~\cite{Arkani-Hamed:2021xlp}~(CE2) and the 1-loop CS-contribution~(CS1)~\cite{deGiorgi:2022xhr}. As it can be seen, the CS1 is negligible, while the CE2 can account at most for $\mathcal{O}(20\%)$ and only for larger masses.  Therefore we keep, for simplicity, only the CE contribution at 1-loop. 

{\it The $(g-2)_\mu$ may receive further contributions from the ALP in all models, but in model A, it appears at NLO and is chirally suppressed.} This can be understood by noticing that the corresponding 1-loop Feynman diagram needs necessarily to contain twice the coupling in Eq.~\eqref{eq:ALP-psi-mu-int}, with the ALP and the $\widehat\psi^-_L$ running into the loop, and thus the chirality flip needs to be in the muon external leg. Naively, we thus expect 
\begin{equation}
    \delta a_\mu^\text{naive}\propto \dfrac{(m_\mu^\text{exp})^2m_R^2}{\alpha_\psi^2\,f_a^4}\,.
\end{equation}
The explicit computation confirms this estimation and reads 
\be
\delta a_\mu^a =\frac{\delta_{|y|,1}}{96\pi^2}\frac{(m_\mu^\text{exp})^2 m_R^2}{f_a^2 M_\psi^2}\,,
\label{ALPg2}
\ee
expressed in terms of the physical mass scale $M_\psi$. With this result, we can neglect the ALP contribution, as we did for the HNL CS1 and CE2 contributions.

Similarly, the radial component $\rho$ also contributes to $\delta a_\mu$. Given that its couplings are very similar to the ones of the ALP, its contribution owns the same properties of Eq.~\eqref{ALPg2} and reads
\be
\delta a_\mu^\rho =\frac{\delta_{|y|,1}}{96\pi^2}\frac{(m_\mu^\text{exp})^2 m_R^2}{f_a^2 m_\rho^2}\times\frac{2+3x_{\rho\psi}\left( 1+\log{x_{\rho\psi}^2}\right)-6x_{\rho\psi}^2+x_{\rho\psi}^3}{(1-x_{\rho\psi})^4}
\ee
where $x_{\rho\psi}\equiv M_\psi^2/m_\rho^2$. This contribution is also negligible with respect to the one in Eq.~\eqref{g2muCE1LNLO}. 

\begin{figure}[h!]
    \centering
     \subfigure[{}\label{fig:g2-dependence}]{\includegraphics[width=0.48\linewidth]{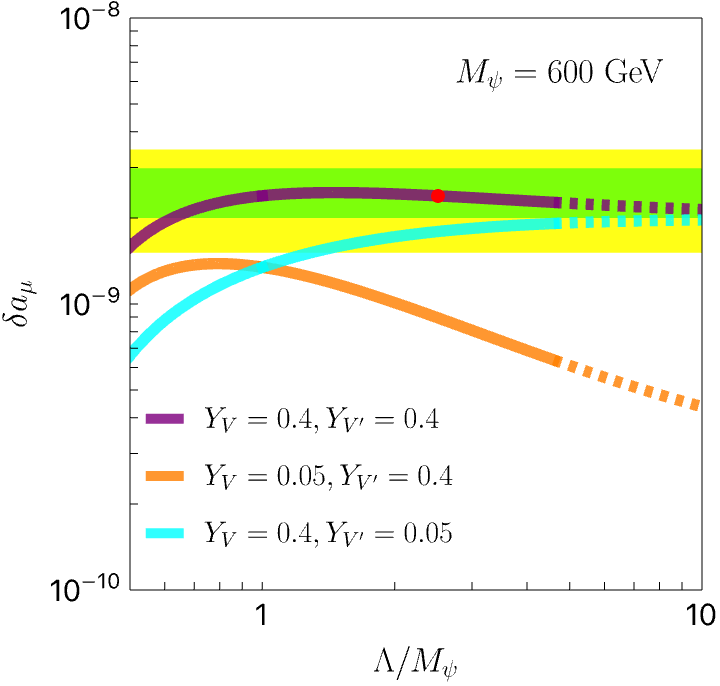}}
     \,
     \subfigure[{}\label{fig:g2-model}]{\includegraphics[width=0.48\linewidth]{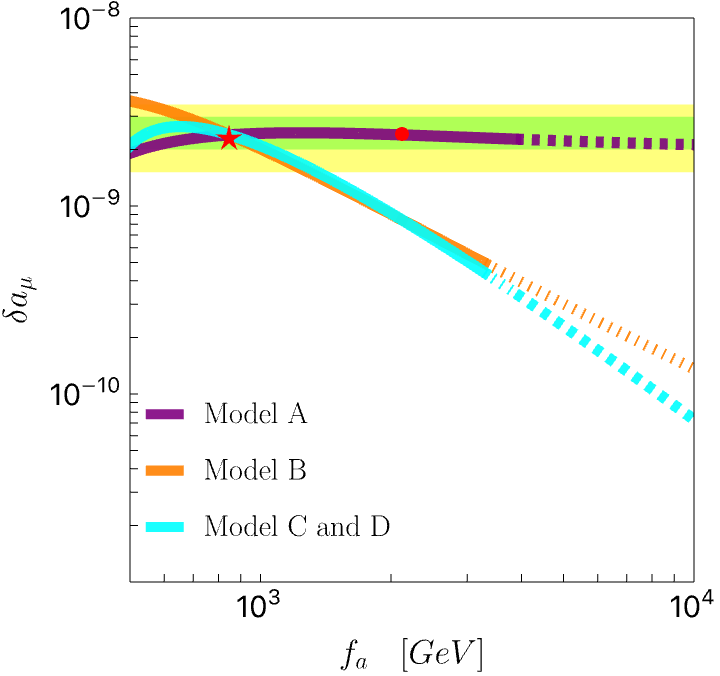}}
    \caption{\em Dependence of $\delta a_\mu $ on the model parameters. $\delta a_\mu^\text{exp}$ values at $1\sigma$ ($2\sigma$) are shown in green (yellow). \textbf{Left:} $\delta a_\mu$ as a function of $\Lambda/M_\psi$ with $M_\psi =600 \GeV$ for $Y_V=Y_{V^\prime}=0.4$ (purple), $Y_{V}=0.05,Y_{V^\prime}=0.4$ (orange) and $Y_{V}=0.4, Y_{V^\prime}=0.05$ (cyan), for any model. $m_N/\Lambda$ and $m_R/M_\psi$ saturate Eqs.~\eqref{eq:ratioN} and \eqref{eq:ratioR}, respectively. The solid (dashed) lines represent the values of $\delta a_\mu$ for  $\abs{Y_N}\leq 1\left(1<\abs{Y_N}\leq 5\right)$. The red dot is a benchmark point of coordinates $\left(\Lambda,M_\psi\right)=(1500,600)\GeV$. \textbf{Right:} $\delta a_\mu$ as a function of the PQ-scale $f_a$ with $Y_V=Y_{V^\prime}=0.4$ for different models: A (purple), B (orange) and C and D (cyan). For simplicity, $\alpha_{N,\psi}$ have been set to 1 according to each model and the remaining free mass, when present, is fixed to $600\GeV$. When the masses are dynamically generated, $\Lambda(M_\psi)=f_a/\sqrt{2}$.  The solid (dashed) lines represent the values of $\delta a_\mu$ for  $\abs{Y_{N,R}}\leq 1\left(1<\abs{Y_{N,R}}\leq 5\right)$. The red dot (star) is a benchmark point of coordinates $\left(\Lambda,M_\psi\right)=(1500,600)\GeV$ for model A (B).}
\label{fig:ycomparisonatlas}
\end{figure}

All in all, the relevant expression for $\delta a_\mu$ is the one given in Eq.~\eqref{g2muCE1LNLO}. Fig.~\ref{fig:ycomparisonatlas} shows the dependence of $\delta a_\mu$ on the model parameters and the type of model realisation.

We first fix $M_\psi=600$ GeV and explore what happens for $Y_V\neq Y_{V^\prime}$ (see Fig.~\ref{fig:g2-dependence}). Such a value of $M_\psi$ is chosen so that the anomaly can be solved for some parameter space. Although for some higher values of $\Lambda/M_\psi$, the case of $Y_V=0.4, Y_{V^\prime}=0.05$ can solve the anomaly at the $2\sigma$ level, in general, a contribution from both $Y_V$ and $Y_{V^\prime}$ ought to be present. Notice that the case with $Y_V=0.05$ does not allow to reach the $2\sigma$ level solution. The red benchmark point is determined by the ratio $\Lambda/M_\psi=2.5$, and slight variations in $\Lambda$ and $M_\psi$ would cause it to shift very slightly to nearby positions on the plot.

On the other hand, the ratio $\Lambda/M_\psi$ is not completely free and depends on the specific model considered, with the exception of models C and D which have the same prediction. In the plot on Fig.~\ref{fig:g2-model}, we are interested in seeing the contribution to the $(g-2)_\mu$ in the different models. Model A is the one that solves the anomaly at $1 \sigma$ for the full parameter range studied. Model B, C and D only solve it for very low values of $f_a$. This plot is obtained fixing for simplicity $\alpha_{N,\psi}=1$ when the two scales $\Lambda$ and/or $M_\psi$ are dynamically generated, while $\Lambda,\,M_\psi=600\GeV$ otherwise, and setting $Y_V=Y_{V^\prime}=0.4$. Choosing higher (lower) values of $Y_{V,V^\prime}$ (not necessarily respecting the equality between both terms), the corresponding line of each model gets shifted towards higher (lower) values of $\delta a_\mu$.

\begin{figure}[h!]
\centering\includegraphics[width=0.6\linewidth]{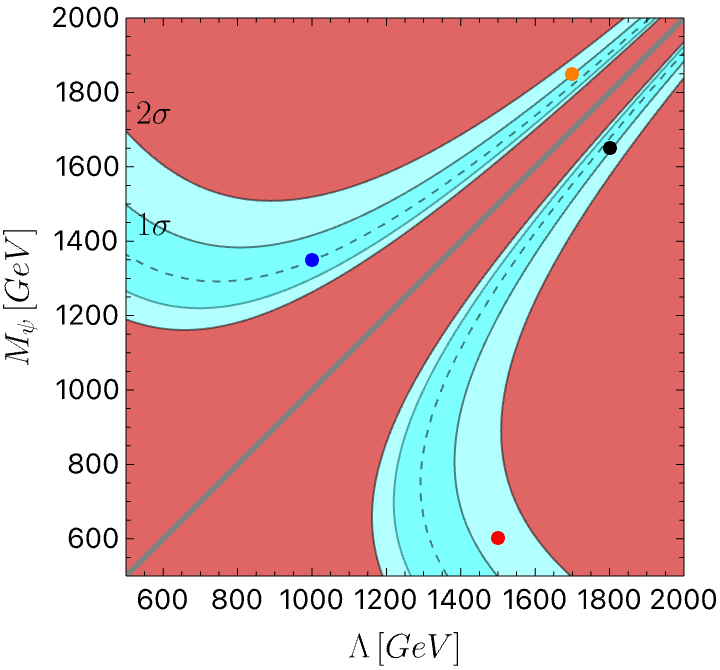}
\caption{\em Regions of the parameter space $\Lambda$ \vs $M_\psi$ to solve the $(g-2)_\mu$ tension for $\delta m_\mu/m_\mu^{\text{exp}}=1$. The dark (light) cyan area corresponds to the parameter space in which $\delta a_\mu$ is $1\sigma\;(2 \sigma)$ of the experimental central value. The red area represents the rest of the values. The benchmark points chosen are $(\Lambda,M_\psi)=(1500,600)\GeV$ (Red), $(1000,1350)\GeV$ (Blue), $(1700,1850)\GeV$ (Orange) and $(1800, 1650)\GeV$ (Black). The grey oblique line shows where qualitatively muon mass vanishes and $Y_{V,V'}$ are thus too large.}
\label{fig:paramatlas}
\end{figure}

Given the similarity of the expression for the muon mass in Eq.~\eqref{deltammu} and the one for the $\delta a_\mu$ in Eq.~\eqref{g2muCE1LNLO}, we can study the correlation between the two observables. For all the models, the ratio of the two expressions reads
\be
\dfrac{\delta a_\mu}{\widehat{m}_\mu}=-\dfrac{6\,m^\text{exp}_\mu\,M_W^2}{\Lambda^2 M_\psi^2}F_0\left(\dfrac{\Lambda^2}{M_W^2},\dfrac{M_\psi^2}{M_W^2}\right)
\left[1+\dfrac{1}{M_\psi^2-\Lambda^2}\left(M^2_\psi
    \log{\dfrac{\mu_\text{R}^2}{M_\psi^2}}-
    \Lambda^2
    \log{\dfrac{\mu_\text{R}^2}{\Lambda^2}}\right)\right]^{-1}\,.
\label{Ratioamumu}
\ee
It is interesting to study the dependence of $\delta a_\mu$ on the parameter space $(\Lambda,\,M_\psi)$ once the muon mass agrees with its experimental determination. We show the results of this analysis in Fig.~\ref{fig:paramatlas}, where we simply take $\widehat{m}_\mu=m^\text{exp}_\mu$ at its central value, given the high precision of the experimental determination. The dark (light) cyan area corresponds to the parameter space in which $\delta a_\mu$ is $1\sigma\;(2 \sigma)$ of the experimental central value. The central grey line corresponds to points where the muon mass is exactly vanishing at 1-loop: for these points, the correlations in Eq.~\eqref{Ratioamumu} are not meaningful. The same applies to the points located near the grey line: in this case, the values of the Yukawa couplings may enter the non-perturbative regime. We will be back on this aspect in the next section.

Although the expression in Eq.~\eqref{Ratioamumu} is independent of the Yukawa couplings $Y_N$, $Y_R$, $Y_V$ and $Y_{V'}$, any point in the plot in Fig.~\ref{fig:paramatlas} corresponds to specific values of the Yukawas: this follows from the condition $\widehat{m}_\mu=m^\text{exp}_\mu$. We selected four benchmark points corresponding to the coloured circles in the plot, with the red circle corresponding to the red circle in Fig.~\ref{fig:g2-model}. The values of the four benchmark points are reported in Tab.~\ref{table:yukawabenchmark}: notice that the Yukawa couplings are not uniquely determined, but there is a flat direction between $Y_V$ and $Y_{V'}$.

\begin{table}[h!]
	\centering
	\begin{tabular}{c| c|c|c| c}
&$\left(\Lambda,M_\psi\right)$~[GeV]&$Y_N$&$Y_R$ &$\abs{Y_V+Y_{V^\prime} M_\psi/\Lambda }$\\
		\hline 
		Red & $\left(1500,600\right)$&0.55&0.26&0.42\\
Blue & $\left(1000,1350\right)$&0.36&0.58&2.26\\
Orange & $\left(1700,1850\right)$&0.62&0.80&2.58\\
Black & $\left(1800,1650\right)$&0.66&0.71&2.23\\
	\end{tabular}
	\caption{\em Summary of the values of the parameters for the different benchmark points shown in Figs.~\ref{fig:ycomparisonatlas} and \ref{fig:paramatlas}. The values of $Y_N$ and $Y_R$ saturate the bounds in Eqs.~\eqref{eq:ratioN} and \eqref{eq:ratioR}.} 
\label{table:yukawabenchmark}
\end{table}

%
%%%%%%%%%%%%%%%%%%%%%%%%%%%%%%%%%%%%%%%%%%%%%%%%%%%%%%%%%%%
%
\subsection{ALP constraints}

The ALP Lagrangian in Eq.~\eqref{ALPLAGChiFlip} describes ALP couplings with gauge bosons and with muons that can be tested experimentally. Recall that bounds on ALP-gauge boson couplings can be placed only when $\delta_{y,1}=1$, while ALP-muons ones can be constrained only when $\ov{\delta}_{x,1}+\ov{\delta}_{y,1}\neq0$ and $Y_V\neq0$. 

\begin{figure}[h!]
\centering
\includegraphics[width=0.9\linewidth]{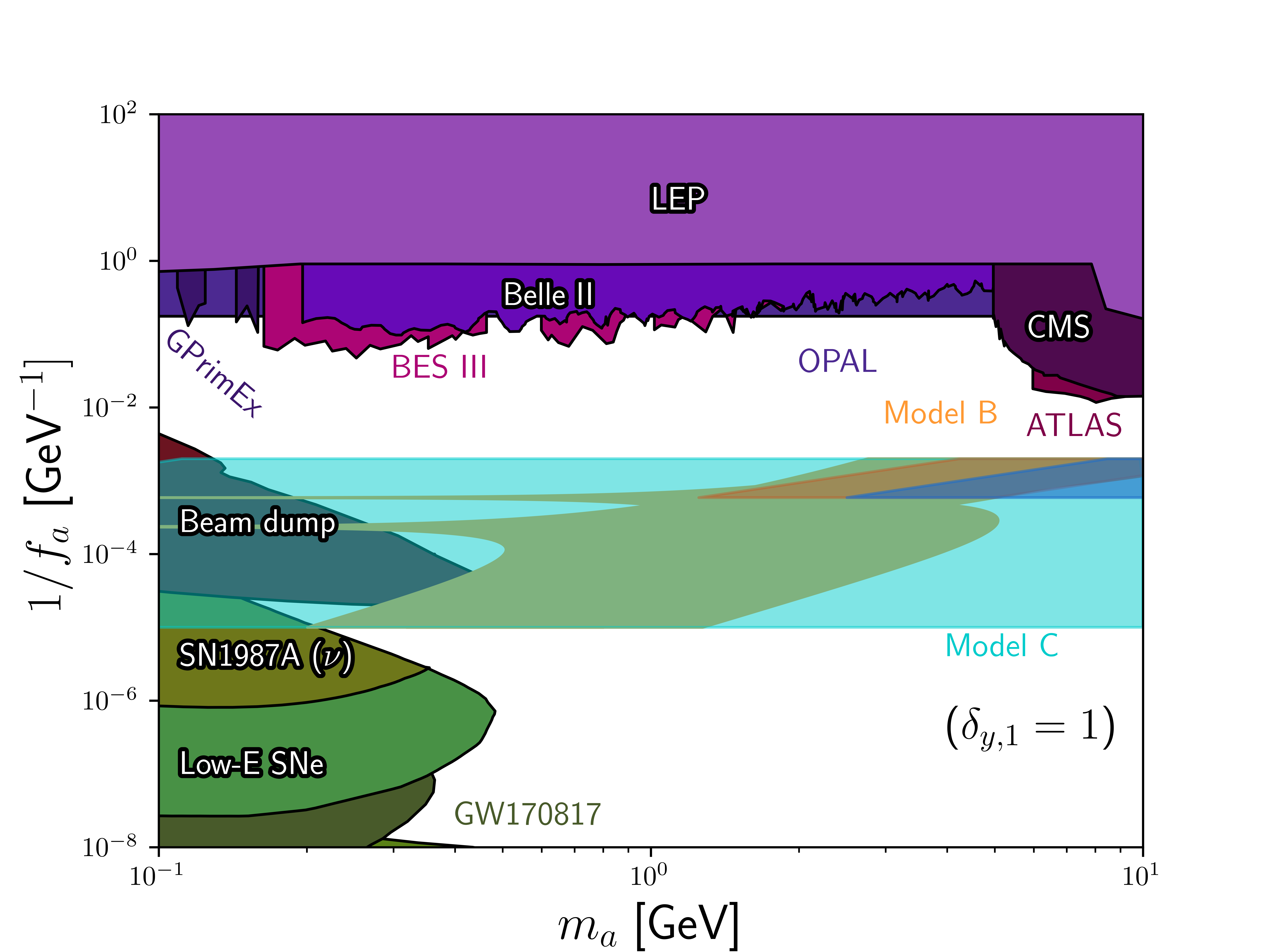}
\caption[]{\em Photon-ALP coupling as a function of $m_a$. Adapted from Ref.~\cite{AxionLimits}. The light orange (cyan) region corresponds to the parameter space for model B (C) when $Y_{V,V^\prime}\subset\left[0.05,0.4\right]$ and $\alpha_{N,\psi}\subset\left[0.5,1.25\right]$. In model B, $\Lambda=1500\GeV$. The darker orange (blue) region, instead, to the benchmark point for model B (C), defined by $\Lambda=1500$, $M_\psi=600\GeV$ and the Yukawas $Y_{V,V^\prime}\subset\left[0.05,0.4\right]$, while $\alpha_{N,\psi}>0.5$.}
\label{fig:photonALP}
\end{figure}

The ALP-photon coupling has been deeply studied and its parameter space is constrained by results coming from detector studies~\cite{ATLAS:2020hii,Dolan:2017osp,CHARM:1985anb,Riordan:1987aw,Blumlein:1990ay,NA64:2020qwq,Belle-II:2020jti,BESIII:2022rzz,CMS:2018erd,Jaeckel:2015jla,PrimEx:2010fvg,Aloni:2019ruo,Dent:2019ueq}, as well as astrophysical searches~\cite{Lucente:2020whw,Caputo:2021rux,Caputo:2022mah, Diamond:2023cto} and cosmology~\cite{Depta:2020wmr}. A summary plot with all the applicable bounds can be seen in Fig.~\ref{fig:photonALP}, for models B and C where $\delta_{y,1}\neq0$. Such a parameter space cannot be studied for model A as the ALP-photon coupling is vanishing. The light orange (cyan) regions correspond to the parameter space for model B (C) when $Y_{V,V^\prime}\subset \left[0.05,0.4\right]$ and $\alpha_{N,\psi}\subset\left[0.5,1.25\right]$. Moreover, in model B $\Lambda=1500\GeV$, and by choosing higher (lower) values the light orange bands displace to the right (left). Additionally, for both models, for larger (smaller) values of $Y_{V,V^\prime}$, the bands of both model B and C displace to the right (left). The dark orange (blue) band represents the parameter space for the benchmark point present in Figs.~\ref{fig:ycomparisonatlas} and \ref{fig:paramatlas}, defined by $\Lambda=1500\GeV$ and $M_\psi=600\GeV$ for model B (C). The darker orange (blue) region is completely contained in the light orange (cyan) one, as expected. The bounds reported in this plot have been considered assuming the most constraining case with the ALP decaying only into photons, that is $\mathrm{Br}(a\to\gamma\gamma)=1$. As we can see, the regions of the parameters space of the models (cyan and orange in the plot) are essentially unconstrained from these bounds, and switching on different decay channels simultaneously would only weaken these constraints.   
Studies on non-resonant vector boson scattering\cite{Bonilla:2022pxu} constrain the ALP couplings to SM gauge bosons, $g_{aWW},\,g_{aZZ}$, leading to
\begin{align}
\abs{g_{aWW}}\leq3.0\,\textrm{TeV}^{-1},&\qquad  \abs{g_{aZZ}}\leq2.8\,\textrm{TeV}^{-1},
\end{align}
which results in an upper limit for $f_a$
\begin{align}
\abs{f_{a_{WW}}}\geq1.7\,\textrm{GeV},&\qquad  \abs{f_{a_{ZZ}}}\geq1.3\,\textrm{GeV},
\end{align}
for masses of $\cO(1)\GeV$ and models B and C. Also in this case, the bounds have been obtained in the most constraining case, that is assuming that the ALP can have only one viable decay channel at a time. Despite this, these limits are extremely weak.

Finally, it is also possible to place an upper limit on the coupling to muons, Eq.~\eqref{eq:loop-ALP-mu}. Being the Lagrangian not shift-symmetric, a direct comparison with the existing bounds in the literature is not straightforward and no general one-to-one matching exists. Indeed, in traditional models, where the ALP arises as the angular component of a scalar field, its couplings to fermions are described in the common chirality-flipping Lagrangian as in Eq.~\eqref{ALPLAGChiFlip}. On the other side, ALP effective descriptions commonly adopt the chirality-preserving or derivative basis, 
\begin{equation}
    \sL_{\derp a}\supset \frac{\partial_\mu a}{f_a}\left(c_{\ell_L}\ov{\ell_L}\gamma^\mu\ell_L+c_{\mu_R}\ov{\mu_R}\gamma^\mu\mu_R\right)\,,
\label{DerivativeBasis}
\end{equation}
where by construction only shift-invariant couplings can be described. Performing a fermion field ALP-dependent redefinitions, it is possible to connect the two bases, but only for what concerns the shift-invariant couplings. In particular, the transformations
\be
\ell_L\to e^{i\,c_{\ell_L}\,a/f_a}\ell_L\,,\qquad\qquad
\mu_R\to e^{i\,c_{\mu_R}\,a/f_a}\mu_R\,,
\ee
remove the couplings in Eq.~\eqref{DerivativeBasis} and generate contributions, besides to anomalous terms, to the chirality-flipping Lagrangian that, below EWSB, reads
\be
\sL_a\supset \widehat{m}_\mu\,e^{i(c_{\mu_R}-c_{\ell_L})\,a/f_a}\ov{\mu_L} \mu_R+\hc\,.
\ee
The existing bounds on the ALP-muon coupling taken from Ref.~\cite{Bauer:2021mvw} apply to the combination 
\be
g_{a\mu\mu}\equiv\frac{c_{\mu_R}-c_{\ell_L}}{f_a}
\ee
that is exactly the shift-invariant couplings written in the chirality-flipping Lagrangian. To estimate the impact of these  bounds on the ALP-muon coupling in our construction that includes shift-breaking terms, we artificially match Eq.~\eqref{eq:loop-ALP-mu} to $g_{a\mu\mu}$ defined in terms of the couplings in the derivative basis, obtaining
\begin{equation}
    g_{a\mu\mu}
    =\dfrac{c_{a\mu\mu}^R}{f_a\, \delta m_\mu}
    =\frac{(\ov{\delta}_{x,1}+\ov{\delta}_{y,1})}{f_a}\times\left(\frac{Y_V}{Y_V+\left(\frac{M_\psi}{\Lambda}\right) Y_{V'}}\right)\,.
    \label{eq:muon-coupling}
\end{equation}
As the matching is artificial, it does not entitle us to use such a result to make claims on $c_{\mu_R}$ nor $c_\ell$. Moreover, the intrinsic error of this procedure resides in neglecting the effects of the anomalous contributions, that being at the 1-loop level are anyway negligible.

This result deserves some additional comments. In any UV model, the tree-level couplings $c_{\ell_L}$ and $c_{\mu_R}$ of the ALP derivative Lagrangian $\sL_{\derp a}$ coincide with the PQ charges of the lepton fields. Although these charges can be taken to be arbitrary small rational numbers, it is very common that $c_{\ell_L},\,c_{\mu_R}\sim\cO(1)$, thus their difference is also expected to be $\cO(1)$, unless the PQ charge assignment is vectorial and in this case, the difference vanishes. As a result, as far as the PQ symmetry is not broken explicitly, the naive estimation for the ALP-muon coupling is $g_{a\mu\mu}\sim\mathcal{O}(1)/f_a$, strongly correlated with the couplings with gauge bosons. There are two caveats in this discussion:
\begin{enumerate}
    \item[(i)] with the tree-level ALP-fermion coupling vanishing, the loop contribution is dominant and spoils the $\sim\mathcal{O}(1)/f_a$ dependence;
    \item[(ii)] with composite fermions, their mixing with other particles with different PQ charges can induce corrections to the ALP-fermion coupling. The latter thus shows a dependence on the mixing parameter that can drastically alter the $\sim\mathcal{O}(1)/f_a$ relation.
\end{enumerate}
However, in both cases, the naive expectation for the ALP-fermion coupling inherited from such contributions can only be $\ll\mathcal{O}(1)/f_a$: in (i) this is due to the loop suppression itself; while in (ii) because the mixing of SM fermions with BSM particles is strongly constrained to be small due to EW-measurements at LHC (cfr.~Eq.s~\eqref{eq:ratioN}-\eqref{eq:ratioR}). Instead, in our case, {\it the ALP-fermion coupling $g_{a\mu\mu}$ spans continuously over various orders of magnitudes, from $\ll\mathcal{O}(1)/f_a$ to $\mathcal{O}(1)/f_a$, providing a proof-of-concept for the existence of models with this peculiar feature.}
This is due to the PQ explicit breaking and effectively removes the direct correlation with $f_a$ decay constant. Given the nature of this breaking, this result is highly model-dependent and cannot be captured in a generic model-independent effective analysis, where instead the ALP-fermion couplings are completely agnostic and should be taken $\cO(1)/f_a$, according to the t'Hooft naturalness principle. 

\begin{figure}[h!]
\centering
\subfigure[{Model A, $Y_V=0.1$}\label{fig:gto-1}]{\includegraphics[width=0.385\linewidth]{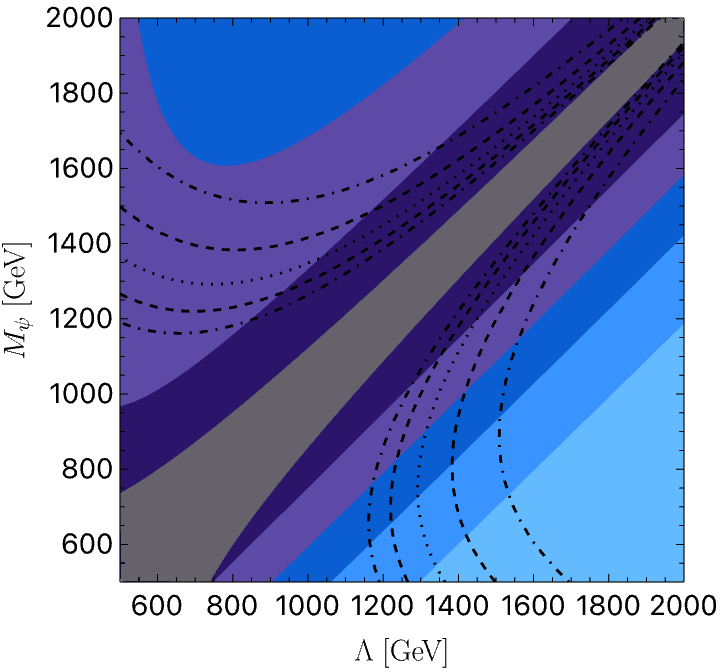}}
\subfigure[{Model A, $Y_V=0.5$}\label{fig:gto-2}]{\includegraphics[width=0.45\linewidth]{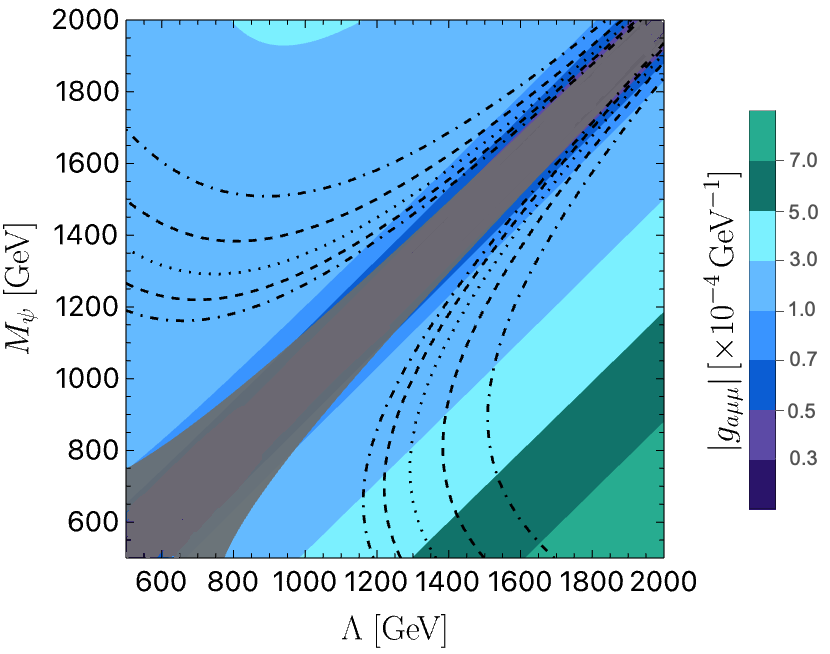}}
\subfigure[{Model C, $Y_V=0.1$}\label{fig:gto-3}]{\includegraphics[width=0.365\linewidth]{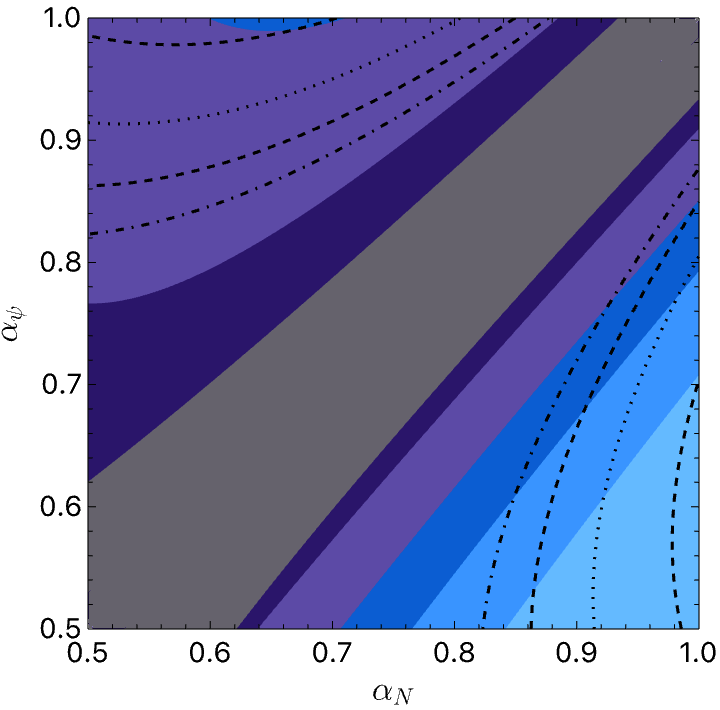}}
\quad\subfigure[{Model C, $Y_V=0.5$}\label{fig:gto-4}]{\includegraphics[width=0.42\linewidth]{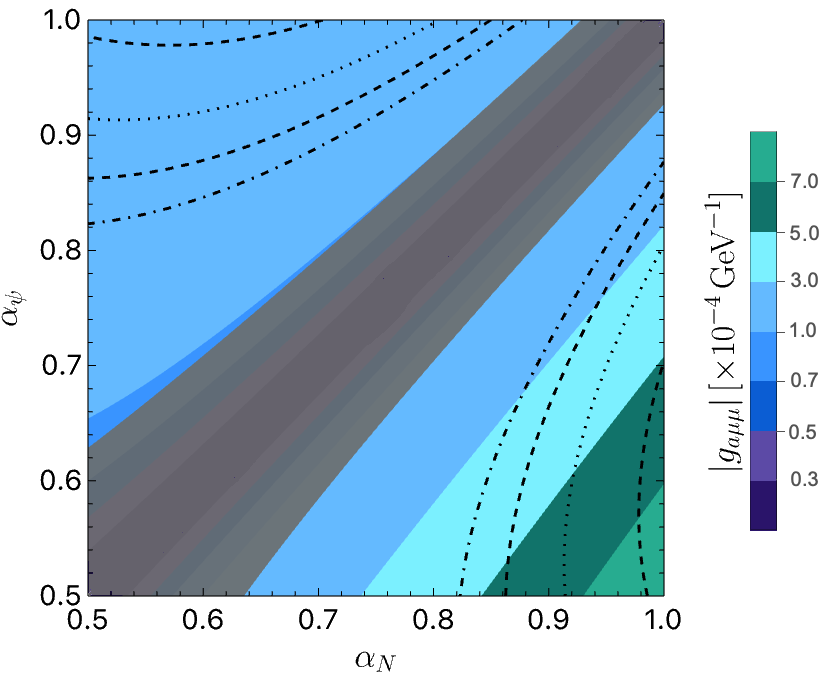}}
\caption[]{\em Coupling of the ALP to muons, $g_{a\mu\mu}$, in relation to the $\left(g-2\right)_\mu$ and the mass of the muon. Coupling values are represented by the different colours in the legend. The parameters are fixed such that $\delta m_\mu=m_\mu^\text{exp}$ in all the parameter space. The dotted(dashed)[dot-dashed] lines show where the central ($1\sigma$)[$2\sigma$] value of $(g-2)_\mu$ can be reproduced.  The grey shaded region stands for the area in which $\abs{Y_{V^\prime}}>5$. 
\textbf{Upper row:} The left (right) plot represents the ALP coupling to muons in the case of model A for values of $Y_V=0.1\left(0.5\right)$. In both cases, $\alpha_N=1$. \textbf{Lower row:} ALP coupling to muons as a function of $\alpha_{N/\psi}$ for model C and $f_a=2\TeV$. In the left (right) plot $Y_V=0.1\left(0.5\right)$.}
\label{fig:gamumucontour}
\end{figure}

In Fig.~\ref{fig:gamumucontour} the relation of the ALP coupling to muons to $\Lambda,\, M_\psi\, \left(\alpha_N,\,\alpha_\psi\right)$ for model A (C) is shown. The results for model B are the same as for model A. Two values of $Y_V$ has been used, namely, $Y_V=0.1 \left(0.5\right)$ for the left (right) plots. The grey band stands for the area in which $\abs{Y_{V^\prime}}>5$, thus breaking perturbativity. In all four plots, the bounds coming from the EW interaction (Eqs.~\eqref{eq:ratioN} and \eqref{eq:ratioR}) do not lead to $Y_{N,R}$ outside of the perturbativity regime.  Additionally, the information on the muon mass and the $(g-2)_\mu$ is placed. Another interesting aspect is that, for model C, the values of $\alpha_{N,\psi}$ that lead to vanishing ALP mass at LO are not allowed. When comparing with the existing literature, we can extract a conservative bound on this coupling from Ref.~\cite{Bauer:2021mvw} (see Fig.~28):
\be
\abs{g_{a\mu\mu}^{\textrm{max}}}=0.01 \GeV^{-1}
\ee
valid in the region $m_a\subset\left[0.1,10\right]\GeV$.
This bound comes from detector searches, specifically from BaBar, where a search for a dark photon in $e^+e^-$ collisions was performed\cite{BaBar:2014zli}.

As we can see, this bound is respected in Fig.~\ref{fig:gamumucontour}.
Stricter bounds could be obtained if the running effects on the coupling of the ALP to the muons were considered. However, this would require a more complex treatment of the theory involving all three generations, which is outside the scope of this work.

\begin{figure}[h!]
\centering
\subfigure[{$\alpha_N=1$}\label{fig:yukdepLambda}]{\includegraphics[width=0.36\linewidth]{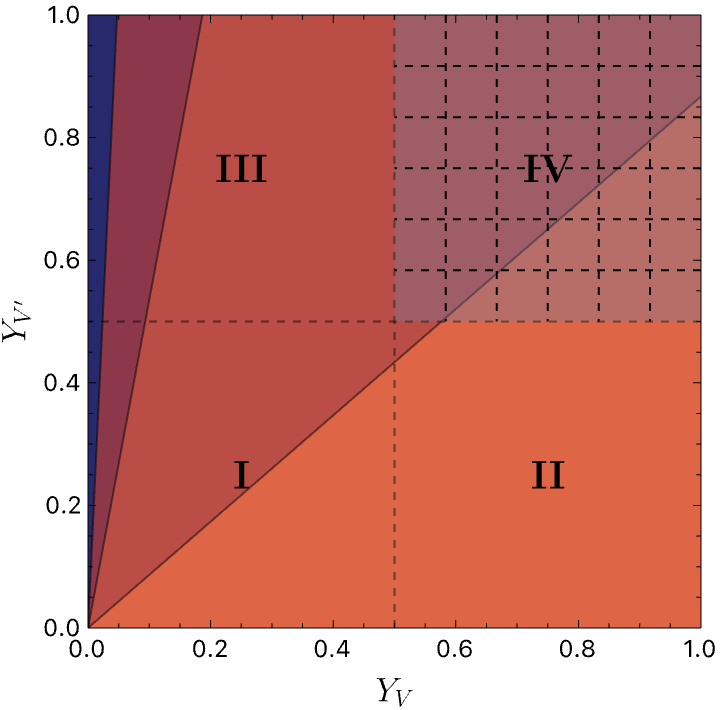}}
\subfigure[{$\alpha_\psi=1$}\label{fig:yukdepMpsi}]{\includegraphics[width=0.45\linewidth]{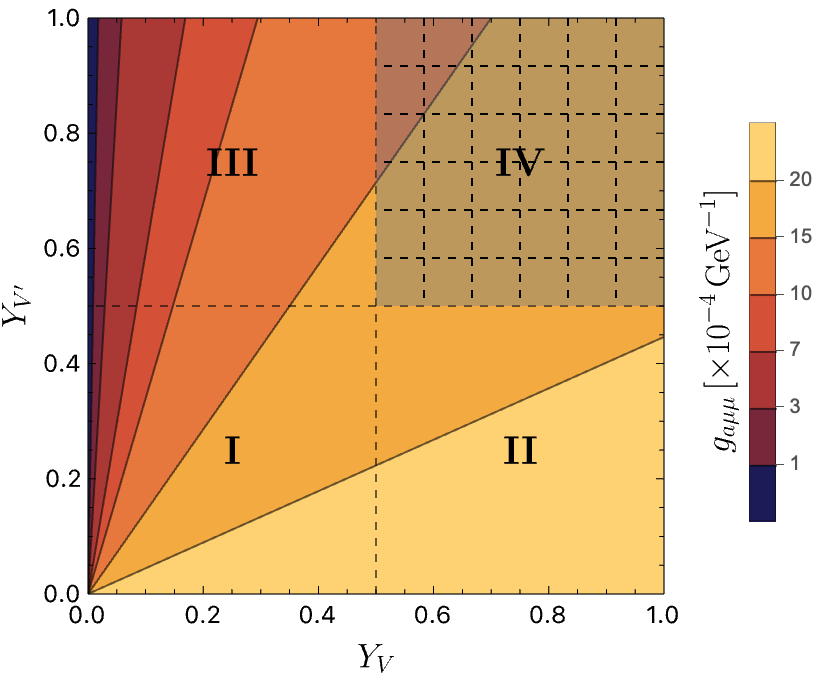}}
\caption[]{\em Coupling $g_{a\mu\mu}$ as a function of $Y_V$, $Y_{V^\prime}$ for model C. The mass scales are taken as $(\Lambda,M_\psi)=(1.5,0.6)\TeV$. The colours refer to the values of $g_{a\mu\mu}$. The parameters are fixed such that $\delta m_\mu=m_\mu^\text{exp}$ in all the parameter space.}
\label{fig:camumuyuk}
\end{figure}

It is also interesting to study the behaviour of the ALP-muon coupling as a function of $Y_V$ and $Y_{V^\prime}$. In Fig.~\ref{fig:camumuyuk}, we take for definiteness model C and fix $(\Lambda,M_\psi)=(1.5,\,0.6)\TeV$, that is the red benchmark point in Figs.~\ref{fig:ycomparisonatlas} and \ref{fig:paramatlas}, for which the $(g-2)_\mu$ tension is solved and the muon mass coincides with its experimental central value. In this case, the $\alpha_{\psi,N}$ parameters are correlated by $\alpha_\psi=\alpha_N\,M_\psi/\Lambda $. The plot in Fig.~\ref{fig:yukdepLambda} is obtained by fixing for simplicity $\alpha_N=1$, while in the plot in Fig.~\ref{fig:yukdepMpsi} $\alpha_\psi=1$. The colours represent the value of the $g_{a\mu\mu}$ coupling: smaller values are in blue while larger values are in yellow, as reported in the side legend. For a given point in this parameter space, the values of the $g_{a\mu\mu}$ coupling in model C are larger than in models A and B by a factor $2$, due to the $(\ov{\delta}_{x,1}+\ov{\delta}_{y,1})$ factor. As we can see in no case the upper bound of $0.01 \GeV^{-1}$ is reached.

Besides showing the distribution of the ALP-muon coupling values in this parameter space, we can comment on the patterns of symmetry breaking. As shown in Eq.~\eqref{PatternSymBreaking}, the two Yukawa couplings $Y_V$ and $Y_{V'}$, when promoted to spurions, have different PQ charges that cannot be both simultaneously vanishing for models A, B and C. When the fermion PQ charges are such that $n_{Y_{V'}}=0$, then the only explicit PQ breaking term is the one associated with $Y_{V}$: we thus expect that $Y_{V}/Y_{V'}\ll1$, corresponding to the top-left region of the plots, labelled as region III. Instead when $n_V=0$, it is the other way around, the term with $Y_{V'}$ is the one breaking the symmetry and thus we expect $Y_{V'}/Y_{V}\ll1$: it corresponds to the bottom-right region, the II one; when both charges are different from zero, the corresponding terms are soft PQ breaking, and $Y_{V'}\sim Y_{V}\ll1$, that is the bottom-left region, the I one. The only side of both plots that is not theoretically viable is the top-right, region IV, where both Yukawa couplings are relatively large: in this case, both of them should be allowed by the symmetry according to the t'Hooft naturalness principle, but this is inconsistent with the spurion charge analysis.

%
%%%%%%%%%%%%%%%%%%%%%%%%%%%%%%%%%%%%%%%%%%%%%%%%%%%%%%%%%%%
%%%%%%%%%%%%%%%%%%%%%%%%%%%%%%%%%%%%%%%%%%%%%%%%%%%%%%%%%%%
%%%%%%%%%%%%%%%%%%%%%%%%%%%%%%%%%%%%%%%%%%%%%%%%%%%%%%%%%%%
%
\section{Conclusions}
\label{sec:Concls}

In contrast to gauge symmetries, our knowledge of global symmetries is very much limited. Within the SM, the combination of Baryon and Lepton numbers $B-L$ is conserved at all quantum orders and, experimentally, the lack of proton decay signals suggests that any violation of the Baryon number, if it occurs at all, is by an exceedingly small amount. For the Lepton number, the situation is different, as indeed the Majorana neutrino option is a key ingredient of many Seesaw realisations. Besides this, the Peccei-Quinn solution to the Strong CP problem involves a spontaneously broken Abelian global symmetry, that needs also to be explicitly broken to equip the axion with a mass. The QCD-axion is essentially a pseudo-Goldstone boson that addresses the Strong CP problem. When it does not fulfil this role, it is often referred to as an ALP. 

In contrast with the QCD-axion case, no universal mechanism to give an ALP a mass has been identified. In this paper, we present a context where a GeV-mass is originated for an ALP decay constant in the range $f_a\subset[0.5,\,100]\TeV$. This is obtained by introducing an additional scalar singlet under the SM gauge symmetries and two lepton $SU(2)_L$-singlets and one vector-like lepton $SU(2)_L$-doublet. This result is novel in the literature, as typically this kind of study is associated with Majorons and the scale $f_a$ is various orders of magnitude larger. We identified three different models, labelled as A, B and C in the main text, where we address different problems: we provide a mass for the ALP and a realistic mass scale for the active neutrinos, solving at the same time the long-standing tension in the anomalous magnetic moment of the muon. Despite being currently subject to strong controversies, the $(g-2)_\mu$ offers a valuable benchmark for BSM phenomenology. If such an anomaly were to fade away, the $(g-2)_\mu$ would become one of the strongest constraints on the model here presented: the ALP phenomenology would quantitatively be different, but qualitatively remains unchanged and thus retains its value.

The peculiar aspect of these models is that the tree-level mass of the muon is forbidden by the global symmetries and it is only generated at $1$-loop. In this way, both the muon mass and the $\delta a_\mu$ have a very similar dependence on the Lagrangian parameters, both receiving a LO contribution at $1$-loop, chirally enhanced due to the presence of the HNLs in the Feynman diagram. We end up with TeV-scale HNLs that may be tested in the future phase of LHC or next-generation colliders. On the other hand, they also participate in the Seesaw mechanism to give realistic masses and mixings to the active neutrinos, via a Linear Seesaw realisation. Besides that, the simultaneous presence of the lepton singlets and doublet allows to give a GeV-mass to the ALP, which we studied both via the CW potential and explicit loop-computations. 

\begin{figure}[h!]
\centering
\subfigure[{Model A, $Y_V=0.1$}\label{fig:to-1}]{\includegraphics[width=0.385\linewidth]{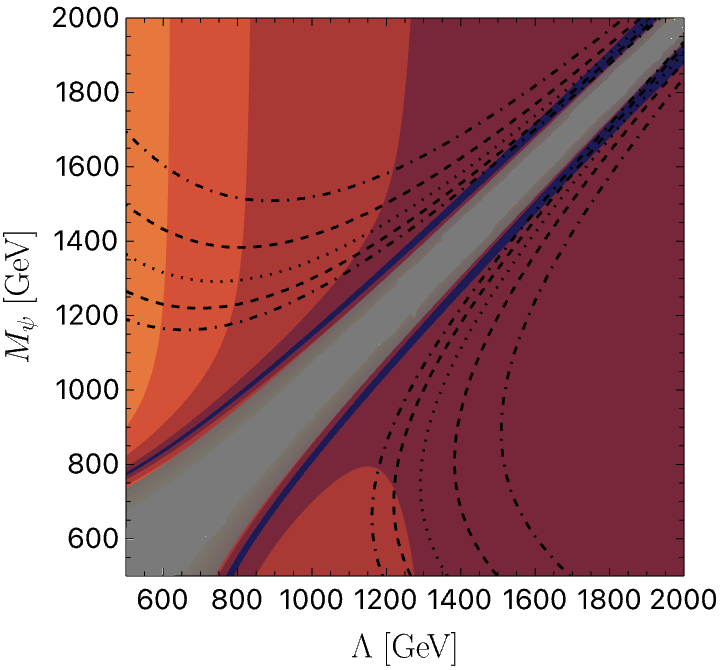}}
\subfigure[{Model A, $Y_V=0.5$}\label{fig:to-2}]{\includegraphics[width=0.45\linewidth]{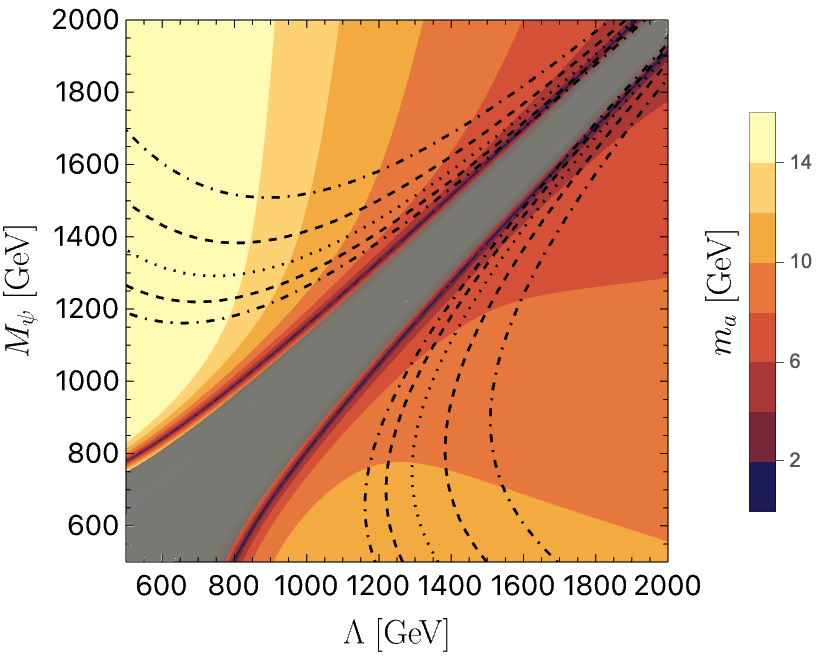}}
\subfigure[{Model C, $Y_V=0.1$}\label{fig:to-3}]{\includegraphics[width=0.35\linewidth]{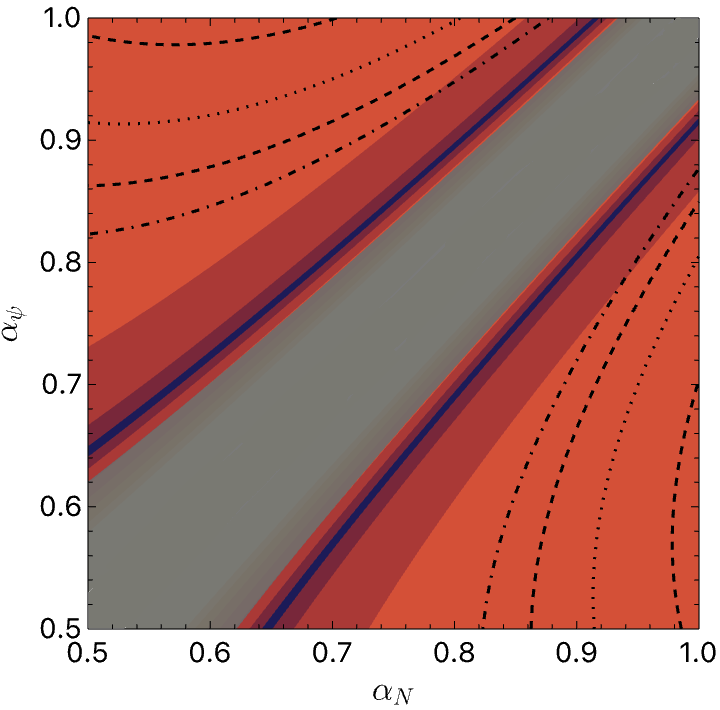}}
\quad\subfigure[{Model C, $Y_V=0.5$}\label{fig:to-4}]{\includegraphics[width=0.42\linewidth]{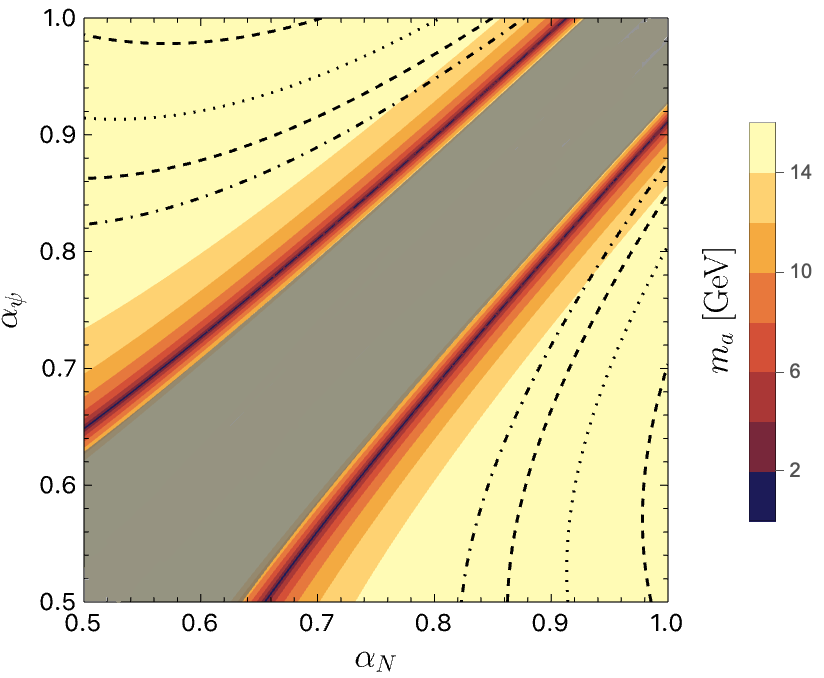}}
\caption[]{\em ALP mass in relation to the $\left(g-2\right)_\mu$ and the mass of the muon. ALP mass values (in GeV) are represented by the different colours in the legend. The dotted line points to the values that explain the experimental muon mass and the central value of the $(g-2)_\mu$. The regions between dashed (dot-dashed) lines explain the experimental muon mass when the $(g-2)_\mu$ is taken to be at  $1\sigma\,\left(2\sigma\right)$ from its central experimental value.   The grey shaded region stands for the area in which $\abs{Y_{V^\prime}}>5$. 
\textbf{Upper row:} The left (right) plot represents the ALP mass in the case of model A for values of $Y_V=0.1\left(0.5\right)$. In both cases, $\alpha_N=1$. \textbf{Lower row:} ALP mass as a function of $\alpha_N/\psi$ for model C and $f_a=2\TeV$. In the left (right) plot $Y_V=0.1\left(0.5\right)$.}
\label{fig:mALPAvsBvsC}
\end{figure}

We can summarise the results on the ALP mass and the $(g-2)_\mu$ tension in Fig.~\ref{fig:mALPAvsBvsC}. Each plot shows the variation of the ALP mass in the $\Lambda$ \vs $M_\psi$ parameter space, that is the HNLs mass scales, overimposing the regions of the $\delta a_\mu$ solution. The plots on the first row pertain to model A, whereas those for model B can be derived from these plots by reflecting them along the diagonal. The ones in the second row, instead, refer to model C. The plots in the first (second) column have been obtained by fixing $Y_V=0.1$ ($Y_V=0.5$). The allowed parameter space, where the active neutrino mass scale is correctly of $\cO(0.1)\eV$, the $(g-2)_\mu$ anomaly is solved and the bounds from EW precision observables and collider direct searches of heavy neutral leptons are satisfied, is pretty large, pinpointing to values for the HNLs masses that cannot be smaller than $1.2\TeV$. Accordingly, the predicted ALP masses are about $\sim1-15\GeV$, depending on the value of $Y_V$. The diagonal grey region for $M_\psi\approx\Lambda$ is theoretically excluded as the muon mass contribution at $1$-loop gets highly suppressed unless the Yukawas compensate acquiring large values, thus entering the non-perturbative regime: the thickness of the grey region is determined by the condition $\abs{Y_{V'}}>5$.

As a specific feature of this model, the ALP-muon coupling spans continuously over various orders of magnitudes, from $\ll\mathcal{O}(1)/f_a$ to $\mathcal{O}(1)/f_a$. This renders our construction a proof-of-concept for the existence of models exhibiting this highly unusual feature: indeed, one would naively expect that the ALP-fermion coupling cannot fluctuate across different orders of magnitude.

Moreover, the ALP-HNLs coupling is an unavoidable byproduct of this construction and can be additionally tested at colliders. As already pointed out in Ref.~\cite{deGiorgi:2022oks}, ALP couplings with exotic fermions may lead to clean observables at hadron colliders. The key ingredient is the fact that tree-level ALP-fermion couplings are proportional to the fermion mass and therefore may be sizable for exotic fermions whose mass has the same origin of the PQ breaking. Ref.~\cite{deGiorgi:2022oks} pointed out that a specific topology called JALZ, which consists in a specific multiple coincident detection of leptons and jets in the final state, would represent a smoking gun to test ALP-HNL couplings. This clearly applies also to our case, although a dedicated analysis would be necessary in order to include into the corresponding analysis the charged-exotic fermions. The aforementioned analysis required ALP-couplings with quarks or gluons, thus non-trivially extending our construction to the quark sector. However, ALP production via photon-fusion or via lepton annihilation, possible in our construction, could be tested with a dedicated analysis at future hadron and muon colliders~\cite{Marcos:2024yfm}.

%
%%%%%%%%%%%%%%%%%%%%%%%%%%%%%%%%%%%%%%%%%%%%%%%%%%%%%%%%%%%%%%%%%%%
%%%%%%%%%%%%%%%%%%%%%%%%%  Acknowledgements    %%%%%%%%%%%%%%%%%%%%
%%%%%%%%%%%%%%%%%%%%%%%%%%%%%%%%%%%%%%%%%%%%%%%%%%%%%%%%%%%%%%%%%%%
%
\section*{Acknowledgements}

The authors would like to express their gratitude to Stefan Pokorski for his valuable contributions during the initial stages of the project.
LM thanks the Department of Physics of the University of California San Diego for hospitality during the development of this project. AdG thanks the University of Heidelberg for its hospitality during the development of this project.
The authors acknowledge partial financial support by the European Union's Horizon 2020 research and innovation programme under the Marie Sk\l odowska-Curie grant agreements No 860881-HIDDeN and 101086085-ASYMMETRY, by the Spanish Research Agency (Agencia Estatal de Investigaci\'on) through the grant IFT Centro de Excelencia Severo Ochoa No CEX2020-001007-S and by the grants PID2019-108892RB-I00 and PID2022-137127NB-I00 funded by MCIN/AEI/ 10.13039/501100011033. The work of AdG is supported by the European Union's Horizon 2020  Marie Sk\l odowska-Curie grant agreement No 860881-HIDDeN. The work of MFZ is supported by the Spanish MIU through the National Program FPU (grant number FPU22/03625). This article is based upon work from COST Action COSMIC WISPers CA21106, supported by COST (European Cooperation in Science and Technology).

%
%%%%%%%%%%%%%%%%%%%%%%%%%%%%%%%%%%%%%%%%%%%%%%%%%%%%%%%%%%%%%%%%%%%
%%%%%%%%%%%%%%%%%%%%%%%%%%%%%  Appendix    %%%%%%%%%%%%%%%%%%%%%%%%
%%%%%%%%%%%%%%%%%%%%%%%%%%%%%%%%%%%%%%%%%%%%%%%%%%%%%%%%%%%%%%%%%%%
%
\newpage
\appendix
\section{Lagrangian in the mass basis}
\label{appendix:Lagrangia-mass-basis}

In this Appendix, we provide the tree-level Lagrangian of Eq.~\eqref{lmodel} in the mass basis, in a generic gauge. We list all interactions with relative corrections of $\mathcal{O}(\Lambda^{-1},M_\psi^{-1})$. As a reference, we adopt the sign conventions of Ref.~\cite{Romao:2012pq}.

The notation for the Higgs doublet components after the EW symmetry breaking reads
\begin{align}
    &H = \begin{pmatrix}
        G^+\\
        \dfrac{v+h+iG_Z}{\sqrt{2}}
    \end{pmatrix}\,, &\widetilde{H} = \begin{pmatrix}
        \dfrac{v+h-iG_Z}{\sqrt{2}}\\
        -G^-
    \end{pmatrix}\,,
\end{align}
where $G^-\equiv (G^+)^\ast$ are the Goldstone bosons and $h$ the physical Higgs particle. 
For the neutral fermions, we use the Majorana spinors defined as
\begin{align}
    &\psi_{1,2}\equiv \psi^0_{L,R}+(\psi^0_{L,R}{})^c\,, &&N\equiv N_R+N_R^c\,, &&S\equiv S_R+S_R^c\,,&&\nu\equiv \nu_L+\nu_L^c\,.
\end{align}
For later convenience, we define the quantities
\be
\begin{gathered}
    s_N \equiv\dfrac{m_N+\epsilon m_S}{\sqrt{2}}\,, \quad
    d_N \equiv\dfrac{m_N-\epsilon m_S}{\sqrt{2}}\,, \quad
    s_V \equiv \dfrac{m_V+m_{V'}}{2}\,,\quad
    d_V \equiv \dfrac{m_V-m_{V'}}{2}\,,\\
    r_+\equiv\dfrac{1}{2}\dfrac{m_V+m_{V'}}{M_\psi-\Lambda}=\dfrac{s_V}{M_\psi-\Lambda}\,, \quad
    r_-\equiv\dfrac{1}{2}\dfrac{m_V-m_{V'}}{M_\psi+\Lambda}=\dfrac{d_V}{M_\psi+\Lambda}\,.
\end{gathered}
\ee

After the tree-level mass diagonalization procedure, the photon-fermion interactions are exactly SM-like, 
\begin{equation}
    \sL_A = e\left(\ov{\widehat{\mu}} \slashed{A}\widehat{\mu}+\ov{\widehat{\psi}^-} \slashed{A}\widehat{\psi}^-\right)\,,
\end{equation}
while the other interactions can be found in Tabs.~\ref{tab:int-h}, \ref{tab:int-Z}, \ref{tab:int-GZ}, \ref{tab:int-W}, \ref{tab:int-GW}, \ref{tab:int-phi}.

\begin{landscape}
\begin{table}[H]
\centering
%\resizebox{1.4\textwidth}{!}{
\begin{tabular}{c|lllllll}
$-\left(\frac{h}{v}\right)\times P_R$  & $\widehat{\mu}$ &$\widehat{\psi}^-$& $\widehat{\nu}$ & $\widehat{N}$ & $\widehat{S} $ & $\widehat{\psi}_1$ & $\widehat{\psi}_2$\\
\toprule
$\ov{\widehat{\mu}}$ & $0$ & $0$ &$0$ &$0$ &$0$ &$0$ &$0$  \\
$\ov{\widehat{\psi}^-}$ & $m_R$ & $\frac{m_R^2}{M_\psi}$ &$0$ &$0$ &$0$ &$0$ &$0$  \\
$\ov{\widehat{\nu}}$ & $0$ &$0$  &  $-2\epsilon\frac{m_N m_S}{\Lambda}$ & $s_N$ & $id_N$ &$(s_Nr_+-d_Nr_-)$ &$i(s_Nr_--d_Nr_+)$\\
&&&&&&$-\frac{s_Ns_V+d_N d_V}{\Lambda}$&$+i\frac{s_Nd_V+d_N s_V}{\Lambda}$\\
$\ov{\widehat{N}}$ & $0$ & $0$ &  $0$ & $\frac{s_N^2}{\Lambda}+\Lambda(r_+^2+r_-^2)$ & $0$ &$s_V$ &$-id_V$\\
&&&&$-M_\psi (r_+^2-r_-^2)$&&&\\
$\ov{\widehat{S}}$ & $0$ & $0$ &  $0$ & $0$ & $\frac{d_N^2}{\Lambda}+\Lambda(r_+^2+r_-^2)$ &$-id_V$ &$-s_V$\\
&&&&&$-M_\psi (r_+^2-r_-^2)$&&\\
$\ov{\widehat{\psi}_1}$ & $0$ & $0$ &  $0$ & $0$ & $0$ &$s_Vr_++d_Vr_-$ &$0$\\
$\ov{\widehat{\psi}_2}$ & $0$ & $0$ &  $0$ & $0$ & $0$ &$0$ &$s_Vr_++d_V r_-$\\
\end{tabular}
%}
\caption{$h$ interactions. All interactions should be multiplied by $-h/v$ and include $P_R$ within the fermionic bilinear. The hermitian conjugate is not included.}
\label{tab:int-h}
\end{table}

\begin{table}[H]
\centering
%\resizebox{1.4\textwidth}{!}{
\begin{tabular}{c|lllllll}
$-\frac{M_Z}{v}Z_\mu\times \gamma^\mu$  & $\widehat{\mu}$ &$\widehat{\psi}^-$& $\widehat{\nu}$ & $\widehat{N}$ & $\widehat{S} $ & $\widehat{\psi}_1$ & $\widehat{\psi}_2$\\
\toprule
$\ov{\widehat{\mu}}$ & $-c_{2\theta_W}P_L+\left[1-c_{2\theta_W}\right]P_R$ & $\frac{m_R}{M_\psi}P_R$ &$0$ &$0$ &$0$ &$0$ &$0$  \\
$\ov{\widehat{\psi}^-}$ & $\frac{m_R}{M_\psi}P_R$ & $-c_{2\theta_W}$ &$0$ &$0$ &$0$ &$0$ &$0$  \\
$\ov{\widehat{\nu}}$ & $0$ &$0$  &  $P_L$ & $\frac{s_N}{\Lambda}P_L$ & $i\frac{d_N}{\Lambda}P_L$ &$0$ &$0$\\
$\ov{\widehat{N}}$ & $0$ & $0$ &  $\frac{s_N}{\Lambda}P_L$ & $0$ & $0$ &$-r_-P_L$ &$ir_+P_L$\\
$\ov{\widehat{S}}$ & $0$ & $0$ &  $-i\frac{d_N}{\Lambda}P_L$ & $0$ & $0$ &$ir_+P_L$ &$r_-P_L$\\
$\ov{\widehat{\psi}_1}$ & $0$ & $0$ &  $0$ & $-r_- P_L$ & $-ir_+ P_L$ &$0$ &$-iP_L$\\
$\ov{\widehat{\psi}_2}$ & $0$ & $0$ &  $0$ & $-ir_+P_L$ & $r_-P_L$ &$iP_L$ &$0$\\
\end{tabular}
%}
\caption{$Z^\mu$ interactions. All interactions should be multiplied by $-M_Z\,Z_\mu/v$ and should include a $\gamma^\mu$ inserted to the left of the projectors. The hermitian conjugate is included.}
\label{tab:int-Z}
\end{table}

\begin{table}[H]
\centering
%\resizebox{1.4\textwidth}{!}{
\begin{tabular}{c|lllllll}
$i\frac{G_Z}{v}\times P_R$  & $\widehat{\mu}$ &$\widehat{\psi}^-$& $\widehat{\nu}$ & $\widehat{N}$ & $\widehat{S} $ & $\widehat{\psi}_1$ & $\widehat{\psi}_2$\\
\toprule
$\ov{\widehat{\mu}}$ & $0$ & $0$ &$0$ &$0$ &$0$ &$0$ &$0$  \\
$\ov{\widehat{\psi}^-}$ & $-m_R$ & $-\frac{m_R^2}{M_\psi}$ &$0$ &$0$ &$0$ &$0$ &$0$  \\
$\ov{\widehat{\nu}}$ & $0$ &$0$  &  $-2\epsilon\frac{m_N m_S}{\Lambda}$ & $s_N$ & $id_N$ &$\sqrt{2}m_S(r_++r_-)$ &$i\sqrt{2}m_S(r_++r_-)$\\
&&&&&&$+\frac{M_\psi}{\Lambda}(s_N r_-+d_N r_+)$&$-i\frac{M_\psi}{\Lambda}(s_N r_++d_N r_-)$\\
$\ov{\widehat{N}}$ & $0$ & $0$ &  $0$ & $\frac{s_N^2}{\Lambda}+2\Lambda r_+ r_-$ & $0$ &$-d_V$ &$is_V$\\
$\ov{\widehat{S}}$ & $0$ & $0$ &  $0$ & $0$ & $\frac{d_N^2}{\Lambda}+2\Lambda r_+ r_-$ &$is_V$ &$d_V$\\
$\ov{\widehat{\psi}_1}$ & $0$ & $0$ &  $0$ & $0$ & $0$ &$-2M_\psi r_+r_-$ &$0$\\
$\ov{\widehat{\psi}_2}$ & $0$ & $0$ &  $0$ & $0$ & $0$ &$0$ &$-2M_\psi r_+ r_-$\\
\end{tabular}
%}
\caption{$G_Z$ interactions. Each term has to be multiplied by $i\,G_Z/v$ and must include a $P_R$ within the fermionic bilinears. The hermitian conjugate is not included.}
\label{tab:int-GZ}
\end{table}

\begin{table}[H]
\centering
%\resizebox{1.4\textwidth}{!}{
\begin{tabular}{c|lllllll}
$-\frac{\sqrt{2}M_W W^-_\mu}{v}$  & $\widehat{\mu}$ &$\widehat{\psi}^-$& $\widehat{\nu}$ & $\widehat{N}$ & $\widehat{S} $ & $\widehat{\psi}_1$ & $\widehat{\psi}_2$\\
\toprule
$\ov{\widehat{\mu}}$ & $0$ & $0$ &$P_L$ &$\frac{s_N}{\Lambda}P_L$ &$i\frac{d_N}{\Lambda}P_L$ &$-\frac{m_R}{\sqrt{2}M_\psi}P_R$ &$i\frac{m_R}{\sqrt{2}M_\psi}P_R$  \\
$\ov{\widehat{\psi}^-}$ & $0$ & $0$ &$0$ &$-\frac{r_++r_-}{\sqrt{2}}P_L-\frac{r_+-r_-}{\sqrt{2}}P_R$ &$-i\frac{r_++r_-}{\sqrt{2}}P_L-i\frac{r_+-r_-}{\sqrt{2}}P_R$ &$\frac{1}{\sqrt{2}}$ &$\frac{-i}{\sqrt{2}}$  \\
$\ov{\widehat{\nu}}$ & $0$ &$0$  &  $0$ & $0$ & $0$ &$0$ &$0$\\
$\ov{\widehat{N}}$ & $0$ & $0$ &  $0$ & $0$ & $0$ &$0$ &$0$\\
$\ov{\widehat{S}}$ & $0$ & $0$ &  $0$ & $0$ & $0$ &$0$ &$0$\\
$\ov{\widehat{\psi}_1}$ & $0$ & $0$ &  $0$ & $0$ & $0$ &$0$ &$0$\\
$\ov{\widehat{\psi}_2}$ & $0$ & $0$ &  $0$ & $0$ & $0$ &$0$ &$0$\\
\end{tabular}
%}
\caption{$W^-_\mu$ interactions. Each term must be multiplied by $-\sqrt{2}M_W\, W^-_\mu/v$ and must include a $\gamma^\mu$ at the left of the projectors in the fermionic bilinear.}
\label{tab:int-W}
\end{table}

\begin{table}[H]
\centering
\resizebox{1.4\textwidth}{!}{
\begin{tabular}{c|lllllll}
$-\left(\frac{\sqrt{2}G^-}{v}\right)$  & $\widehat{\mu}$ &$\widehat{\psi}^-$& $\widehat{\nu}$ & $\widehat{N}$ & $\widehat{S} $ & $\widehat{\psi}_1$ & $\widehat{\psi}_2$\\
\toprule
$\ov{\widehat{\mu}}$ & $0$ & $0$ & $\left[2\epsilon\frac{m_N m_S}{\Lambda}\right]P_R$&$\left[-\frac{m_R}{\sqrt{2}}\frac{\Lambda}{M_\psi}(r_+-r_-)\right]P_L$&$\left[-i\frac{m_R}{\sqrt{2}}\frac{\Lambda}{M_\psi}(r_+-r_-)\right]P_L$&$\left[\frac{m_R}{\sqrt{2}}\right]P_L$&$\left[-i\frac{m_R}{\sqrt{2}}\right]P_L$\\
&&& &$+\left[-s_N\right]P_R$&$+\left[-id_N\right]P_R$&$+\left[d_N r_--s_Nr_+\right]P_R$&$+\left[i(d_N r_+-s_N r_-)\right]P_R$\\
$\ov{\widehat{\psi}^-}$ & $0$ & $0$ &$\left[\frac{m_N m_V}{\Lambda}\right]P_L$&$\left[-\frac{m_V}{\sqrt{2}}\right]P_L$&$\left[-i\frac{m_V}{\sqrt{2}}\right]P_L$&$\left[-\frac{m_V}{\sqrt{2}}(r_++r_-)+\frac{m_R^2}{\sqrt{2}M_\psi}\right]P_L$&$\left[i\frac{m_V}{\sqrt{2}}(r_++r_-)-i\frac{m_R^2}{\sqrt{2}M_\psi}\right]P_L$\\
&&&$+\left[\epsilon \frac{m_S m_{V'}}{\Lambda}\right]P_R$&$+\left[-\frac{m_{V'}}{\sqrt{2}}\right]P_R$&$+\left[-i\frac{m_{V'}}{\sqrt{2}}\right]P_R$&$+\left[-\frac{m_{V'}}{\sqrt{2}}(r_+-r_-)\right]P_R$&$+\left[i\frac{m_{V'}}{\sqrt{2}}(r_+-r_-)\right]P_R$\\
$\ov{\widehat{\nu}}$ & $0$ &$0$  &  $0$ & $0$ & $0$ &$0$ &$0$\\
$\ov{\widehat{N}}$ & $0$ & $0$ &  $0$ & $0$ & $0$ &$0$ &$0$\\
$\ov{\widehat{S}}$ & $0$ & $0$ &  $0$ & $0$ & $0$ &$0$ &$0$\\
$\ov{\widehat{\psi}_1}$ & $0$ & $0$ &  $0$ & $0$ & $0$ &$0$ &$0$\\
$\ov{\widehat{\psi}_2}$ & $0$ & $0$ &  $0$ & $0$ & $0$ &$0$ &$0$\\
\end{tabular}
}
\caption{$G^-$ interactions. Each term has to be multiplied by $-\sqrt{2}G^-/v$. The projectors $P_{L,R}$ must be included within the fermionic bilinears.}
\label{tab:int-GW}
\end{table}

\begin{table}[H]
\centering
%\resizebox{1.4\textwidth}{!}{
\begin{tabular}{c|lllllll}
$-\frac{\sqrt{2}}{f_a}\phi^{(\ast)}\times P_R$  & $\widehat{\mu}$ &$\widehat{\psi}^-$& $\widehat{\nu}$ & $\widehat{N}$ & $\widehat{S} $ & $\widehat{\psi}_1$ & $\widehat{\psi}_2$\\
\toprule
$\ov{\widehat{\mu}}$ & $0$ & $0$ &$0$ &$0$ &$0$ &$0$ &$0$  \\
$\ov{\widehat{\psi}^-}$ & $-\delta_{|y|,1}m_R$ & $\delta_{|y|,1}M_\psi$ &$0$ &$0$ &$0$ &$0$ &$0$  \\
$\ov{\widehat{\nu}}$ & $0$ &$0$  &  $0$ & $-\delta_{|x|,1}s_N$ & $-i\delta_{|x|,1}d_N$ &$0$ &$0$\\
$\ov{\widehat{N}}$ & $0$ & $0$ &  $0$ & $\delta_{|x|,1}\frac{\Lambda}{2}$ & $0$ &$(\delta_{|x|,1}\Lambda-\delta_{|y|,1}M_\psi)r_+$ &$i(\delta_{|x|,1}\Lambda+\delta_{|y|,1}M_\psi)r_-$\\
$\ov{\widehat{S}}$ & $0$ & $0$ &  $0$ & $0$ & $\delta_{|x|,1}\frac{\Lambda}{2}$ &$i(\delta_{|x|,1}\Lambda+\delta_{|y|,1}M_\psi)r_-$ &$-(\delta_{|x|,1}\Lambda-\delta_{|y|,1}M_\psi)r_+$\\
$\ov{\widehat{\psi}_1}$ & $0$ & $0$ &  $0$ & $0$ & $0$ &$\delta_{|y|,1}\frac{M_\psi}{2}$ &$0$\\
$\ov{\widehat{\psi}_2}$ & $0$ & $0$ &  $0$ & $0$ & $0$ &$0$ &$\delta_{|y|,1}\frac{M_\psi}{2}$\\
\end{tabular}
%}
\caption{$\phi$ interactions. Each term has to be multiplied by $-\sqrt{2}\phi^{(\ast)}/f_a$. $P_R$ must be inserted in all fermionic bilinears. The choice of $\phi$ or $\phi^\ast$ depends on the model and should be placed in front of the corresponding $\delta_{|x|,1}$ or $\delta_{|y|,1}$. The hermitian conjugate is not included.}
\label{tab:int-phi}
\end{table}
\end{landscape}

%%%%%%%%%%%%%%%%%%%%%%%%%%%%%%%%%%%%%%%%%%%%%%%%%%%%%%%%%%%%%%%%%%%%%%%%%%%%%%%%%%%%%%%%%%%%%%%%%%%%%%%%%%%%%%%%%%%%%%%%%%%%%%%%%%%%%%
\section{ALP mass full result}
\label{appendix:ALP-full-CW}
The ALP potential can be obtained using the Coleman-Weinberg potential~\cite{Coleman:1973jx} which for Weyl fermions in dimensional regularization $d=4-2\varepsilon$ reads
\be
V_{\text{CW}}=-\frac{1}{2}\cdot\frac{1}{16\pi^2}\left\{\Tr{\left[\left(M_\chi M_\chi^\dagger\right)^2\log\left(\frac{M_\chi M_\chi^\dagger}{\mu_R^2}\right)\right]}-\Tr{\left[\left(M_\chi M_\chi^\dagger\right)^2\right]\left(\frac{3}{2}+\frac{1}{\widetilde{\varepsilon}}\right)}\right\}
\label{cwpotential}
\ee
 where $\mathcal{M}_\chi$ is the $5\times5$ neutral mass matrix of Eq.~\eqref{eq:massmatrix}, $\mu_R$ the renormalization scale, and
\be
\dfrac{1}{\widetilde{\varepsilon}}\equiv \dfrac{1}{\varepsilon}-\gamma_E+\log(4\pi)\,,
\ee
where $\gamma_E$ is the Euler-Mascheroni constant. The charged mass matrix does not contribute to the ALP potential.
By exploiting the properties of the trace, the above expression can be written in terms of the eigenvalues, $\{\lambda_i\}_{i=1}^5$, of $\mathcal{M}_\chi \mathcal{M}_\chi^\dagger$
\be
V_{\text{CW}}=-\frac{1}{2}\cdot\frac{1}{16\pi^2}\sum\limits_{i=1}^{5}\left\{\lambda_i^2\log\left(\frac{\lambda_i^2}{\mu_R^2}\right)-\lambda_i^2\left(\frac{3}{2}+\frac{1}{\widetilde{\varepsilon}}\right)\right\}\,.
\ee 
We can further simplify the problem. One eigenvalue can be safely neglected due to its smallness as it should be of order of the neutrino mass. Furthermore, the remaining four eigenvalues are equal two to two (cfr.~Eq.~\eqref{FinalMassesNeutral}), reducing the problem to the computation of only two eigenvalues here called $\lambda_{1,2}$.
They can be conveniently extracted from the traces of $\mathcal{M}_\chi\mathcal{M}_\chi^\dagger$
\begin{align}
    &\Tr \left(\mathcal{M}_\chi \mathcal{M}_\chi^\dagger\right)\approx 2(\lambda_1+\lambda_2)\,, & \Tr (\mathcal{M}_\chi\mathcal{M}_\chi^\dagger)^2\approx2(\lambda_1^2+\lambda_2^2)\,,
\end{align}
yielding
\begin{equation}
    \lambda_{1,2}\approx\frac{\Tr \left(\mathcal{M}_\chi\mathcal{M}_\chi^\dagger\right) \pm \sqrt{4 \Tr \left(\mathcal{M}_\chi\mathcal{M}_\chi^\dagger\right)^2-\left(\Tr \mathcal{M}_\chi\mathcal{M}_\chi^\dagger\right)^2}}{4}\,.
\end{equation}

Finally, the mass of the ALP can be obtained via
\begin{equation}
    m_a^2 \equiv \left.\frac{\partial^2 V_\text{CW}}{\partial a^2}\right|_{a=0}\,.
\end{equation}
We now turn to the computation of the potential for the four models studied in this paper. The associated mass matrices for each model are
\begin{align}
&\text{\textbf{Model A:}} &&\mathcal{M}_{\chi,A}=\left(
    \begin{array}{ccccc}
         0& m_N& m_S&0&0\\
         m_N& 0&\alpha_N \phi^{(\ast)}&m_V^\prime&0\\
         m_S&\alpha_N \phi^{(\ast)}&0&0&m_V\\
         0&m_V^\prime&0&0&M_\psi\\
         0&0&m_V&M_\psi&0\\
    \end{array}
    \right)\,,\\
    &\text{\textbf{Model B:}} &&\mathcal{M}_{\chi,B}=\left(
    \begin{array}{ccccc}
         0& m_N& m_S&0&0\\
         m_N& 0&\Lambda&m_V^\prime&0\\
         m_S&\Lambda&0&0&m_V\\
         0&m_V^\prime&0&0&\alpha_\psi\phi^{(\ast)}\\
         0&0&m_V&\alpha_\psi\phi^{(\ast)}&0\\
    \end{array}
    \right)\,,
    \end{align}
    \begin{align}
    &\text{\textbf{Model C:}} &&\mathcal{M}_{\chi,C}=\left(
    \begin{array}{ccccc}
         0& m_N& m_S&0&0\\
         m_N& 0&\alpha_N \phi^{(\ast)}&m_V^\prime&0\\
         m_S&\alpha_N \phi^{(\ast)}&0&0&m_V\\
         0&m_V^\prime&0&0&\alpha_\psi\phi^{(\ast)}\\
         0&0&m_V&\alpha_\psi\phi^{(\ast)}&0\\
    \end{array}
    \right)\,,\\
    &\text{\textbf{Model D:}} &&\mathcal{M}_{\chi,D}=\left(
    \begin{array}{ccccc}
         0& m_N& m_S&0&0\\
         m_N& 0&\alpha_N \phi&m_V^\prime&0\\
         m_S&\alpha_N \phi&0&0&m_V\\
         0&m_V^\prime&0&0&\alpha_\psi\phi^\ast\\
         0&0&m_V&\alpha_\psi\phi^\ast&0\\
    \end{array}
    \right)\,.
\end{align}
Notice that in model A, B and C $\alpha_{N/\psi}$ can be proportional to either $\phi$ or $\phi^\ast$, while model D is characterised by having $\alpha_N \phi$, $\alpha_\psi \phi^\ast$ or $\alpha_N \phi^\ast$, $\alpha_\psi \phi$.

By explicit computation one finds
\begin{equation}
    \begin{aligned}
        V_\text{CW}&\supset a^2(\ov{\delta}_{x,1}+\ov{\delta}_{y,1})^2 \frac{m_{V^\prime} m_V \Lambda M_\psi}{8\pi^2 f_a^2}\left\{\frac{1}{2}\left[\frac{c}{d}\log\left(\frac{c+d}{c-d}\right)+\log\left(\frac{c^2-d^2}{\mu_R^4}\right)\right]-1-\frac{1}{\widetilde{\varepsilon}}\right\}\,,\\
        m_a^2&=(\ov{\delta}_{x,1}+\ov{\delta}_{y,1})^2 \frac{m_{V^\prime} m_V \Lambda M_\psi}{4\pi^2 f_a^2}\left\{\frac{1}{2}\left[\frac{c}{d}\log\left(\frac{c+d}{c-d}\right)+\log\left(\frac{c^2-d^2}{\mu_R^4}\right)\right]-1-\frac{1}{\widetilde{\varepsilon}}\right\}\,,\\
        c&=\frac{1}{2}\left(m_N^2+m_V^2+m_{V^\prime}^2+\Lambda^2+M_\psi^2\right)\,,\\
        d&=\sqrt{c^2-m_N^2\left(m_V^2+M_\psi^2\right)-m_{V^\prime}^2 m_V^2-\Lambda^2 M_\psi^2+2 m_V m_{V^\prime} \Lambda M_\psi}\,,
    \end{aligned}
    \label{CWgeneral}
\end{equation}
where $(\ov{\delta}_{x,1},\ov{\delta}_{y,1})$ are two parameters that interpolate among all models and follow the convention established in the main text.
Taking the limit of large $\Lambda$, $M_\psi$, leads to the LO expression of Eq.~\eqref{eq:massALPatLO}.

%%%%%%%%%%%%%%%%%%%%%%%%%%%
\section{Details on the ALP coupling to muons}
\label{ALPMuonsCouplings}
In this Appendix, we further investigate the result of the ALP-muon coupling obtained in Eq.~\eqref{eq:loop-ALP-mu}. This result might at first sight look obscure as it comes from a combination of tree-level mixing and loop results. We show here how from symmetry arguments the situation is more transparent and how the result can be obtained solely from them.

As can be noticed from Eq.~\eqref{eq:loop-ALP-mu}, the LO ALP coupling to on-shell muons is zero if
\begin{equation}
    c_{a\mu\mu}^R=-c_{a\mu\mu}^L = 0 \qquad \Leftrightarrow \qquad Y_V=0\,\vee \ov{\delta}_{x,1}=-\ov{\delta}_{y,1}\,,
\end{equation}
and otherwise closely resembles the 1-loop muon mass. Such features, which might seem peculiar at first glance, can be understood on more general grounds. To appreciate it, let us first consider the Lagrangian to be PQ-invariant, and let us assume that, still in the interaction-basis, we rotate away the ALP from the Yukawa sector via the rotation
\begin{equation}
    \Psi \to e^{i\chi_\Psi\frac{a}{f_a}}\Psi\,,
\end{equation}
being $\chi_\Psi$ the PQ-charge of the generic fermion $\Psi$. As the symmetry is assumed to be exact, the ALP Lagrangian will only contain interactions of the form
\begin{equation}
\label{eq:derivative-charged}
    %\begin{split}
        \sL_a 
        %&\supset -\frac{\partial_\mu a}{f_a}\left(\sum\limits_i \chi_{\Psi_i}\ov{\Psi_i}\gamma^\mu \Psi^i\right)\,,\\
        %&
        \supset -\frac{\partial_\mu a}{f_a}\left( \chi_{\mu_L}\ov{\mu_L}\gamma^\mu \mu_L+\chi_{\mu_R}\ov{\mu_R}\gamma^\mu \mu_R+\chi_{\psi_L^-}\ov{\psi^-_L}\gamma^\mu \psi^-_L+\chi_{\psi^-_R}\ov{\psi^-_R}\gamma^\mu \psi^-_R\right)\,.
    %\end{split}
\end{equation}
To go to the mass basis, at tree-level, we just need to rotate the right-handed fields (cfr.~\eqref{eq:charged-mixing}). We are interested in the coupling of the ALP to muons in the chirality-flipping basis. When considering on-shell fermions, one can move from the derivative- to the chirality-flipping basis integrating by parts and applying the equations of motion of the fermions. Using such a procedure, the couplings of the ALP to fermions are found to be proportional to their masses~(see e.g. Ref.~\cite{Bonilla:2022qgm}). As in our case the mass of the muon is generated at 1-loop, all 1-loop contributions to the derivative couplings of Eq.~\eqref{eq:derivative-charged} formally become 2-loops when moving to the chirality-flipping basis, and thus must be neglected accordingly to the approximation we took throughout the paper. All in all, this implies that at 1-loop it is sufficient to consider only the tree-level mixing of the fields, yielding
\begin{equation}
\label{eq:ALP-muon-lag}
    \begin{split}
        \sL_a 
        &\supset -\frac{\partial_\mu a}{f_a}\left[ \chi_{\mu_L}\ov{\widehat{\mu}}\gamma^\mu \widehat{\mu}+\left(\left(\chi_{\mu_R}-\chi_{\mu_L}\right)-\left(\chi_{\mu_R}-\chi_{\psi_R}\right)\left(\frac{m_R^2}{M_\psi^2}\frac{1}{1+m_R^2/M_\psi^2}\right)\right)\ov{\widehat{\mu}_R}\gamma^\mu \widehat{\mu}_R\right]\,.
    \end{split}
\end{equation}
The first term vanishes when applying the equations of motion, while the second generates the term
\begin{equation}
\label{eq:ALP-muon-expected}
    \sL_a\supset \frac{a}{f_a}\left[\left(\chi_{\mu_R}-\chi_{\mu_L}\right)-\left(\chi_{\mu_R}-\chi_{\psi_R}\right)\left(\frac{m_R^2}{M_\psi^2}\frac{1}{1+m_R^2/M_\psi^2}\right)\right]\,\underbrace{\widehat{m}_\mu}_{=\delta m_\mu}\,\ov{\widehat{\mu}}i\gamma_5\widehat{\mu} \,.
\end{equation}
This result, valid assuming an exact PQ-symmetry, has two main consequences:
\begin{enumerate}
    \item the coupling of the muon to the ALP must be proportional to its mass weighted by the PQ-charges;
    \item the coupling at LO vanishes if $\chi_{\mu_L}=\chi_{\mu_R}$, i.e. if the rotation is vectorial, and is not zero but suppressed by an extra $m_R^2/M_\psi^2$ if and only if $\chi_{\mu_L}\neq\chi_{\mu_R}$.
\end{enumerate}

Let us now examine the result of Eq.~\eqref{eq:loop-ALP-mu} under such considerations. In our case, the PQ-symmetry is not exact, so we have to take some extra care. As we learnt from the computation of $m_a^2$, the PQ is restored if $Y_V=0$, $Y_{V'}=0$, or in model D independently of $Y_{V,V'}$. Let us examine the different cases separately to check the correctness of the consistency of the computation.

\paragraph{$\mathbf{Y_V=0}$.} In this case, one can see that $\chi_{\mu_L}=\chi_{\mu_R}$ is a possible charge assignment in all the models, meaning that we expect no ALP-muon coupling at LO. This is consistent with the explicit computation of Eq.~\eqref{eq:loop-ALP-mu}.

\paragraph{$\mathbf{Y_{V'}=0}$.}
In this case, the coupling computed in Eq.~\eqref{eq:loop-ALP-mu} is exactly proportional to the muon mass.
The charges in the different models are given in Tab.~\ref{tab:charges-YV'0}. The case of model D is discussed in the next section, as it is independent of $Y_{V'}$ being zero or not. As it can be seen from the charges, we always find that $\ov{\delta}_{x,1}+\ov{\delta}_{y,1}=(\chi_{\mu_R}-\chi_{\mu_L})$, confirming the matching between explicit loop calculation and the expected result of Eq.~\eqref{eq:ALP-muon-expected}.

\begin{table}[h!]
    \centering
    \begin{tabular}{c|rrrrr}
    \toprule
        Model &$\chi_{\mu_L}$ &$\chi_{\mu_R}$& $\chi_{\psi_R}$&$\ov{\delta}_{x,1}$ & $\ov{\delta}_{y,1}$\\
    \midrule
         A&$-1/2$&$1/2$&$1/2$&$1$&$0$  \\
         B&$0$&$-1$&$0$&$0$&$-1$ \\
         C&$-1/2$& $3/2$&$1/2$&$1$&$1$ \\
         D&$-1/2$&$-1/2$&$1/2$&$1$&$-1$ \\
    \bottomrule
    \end{tabular}
    \caption{\em PQ-charges of the fields in the case $Y_{V'}=0$ for the different models.}
    \label{tab:charges-YV'0}
\end{table}

\paragraph{Model D.}
In this case, one finds that $\chi_{\mu_L}=\chi_{\mu_R}=-1/2$, and so the LO coupling is zero. Almost magically, in this case $\ov{\delta}_{x,1}=-\ov{\delta}_{y,1}$, setting to zero the explicitly computed 1-loop ALP-muon coupling. 

Despite the PQ being explicitly broken, one can still artificially match the result to the standard ALP EFT in the derivative basis
\begin{equation}
    \sL_a^\text{EFT}\supset \frac{\partial_\mu a}{f_a}\left(c_{\ell_L}\ov{\ell_L}\gamma^\mu\ell_L+c_{\mu_R}\ov{\mu_R}\gamma^\mu\mu_R\right)\,,
\end{equation}
by imposing
\begin{equation}
    \frac{c_{\mu_R}-c_{\ell_L}}{f_a}
    =\dfrac{c_{a\mu\mu}^R}{f_a\, \delta m_\mu}
    =\frac{(\ov{\delta}_{x_1}+\ov{\delta}_{y,1})}{f_a}\times\left(\frac{Y_V}{Y_V+\left(\frac{M_\psi}{\Lambda}\right) Y_{V'}}\right)\,.
%    \label{eq:muon-coupling}
\end{equation}
The only parameters that appear in this result are the same ones that appear also in the computation of $m_a^2$; this is no coincidence as they are the only ones responsible for the (in)exactness of the PQ-symmetry.
If the PQ-symmetry was exact the coupling would only depend on the PQ-charges of the muon; in the case of broken PQ it just retains a proportionality to them, allowing for arbitrarily small values of $c_{\ell_L}-c_{\mu_R}$ depending on $Y_{V,V'},\,\Lambda,\,M_\psi$. This fact seems trivial from the starting point of the ALP EFT where the Wilson coefficients do not necessarily play the role of PQ-charges and can thus pick any values, but it is highly non-trivial from the point of view of the UV theory.
Interestingly, this coupling can be made arbitrarily small by taking $Y_V\to 0$ and still reproducing the correct muon mass via $Y_{V'}$.

Remarkably, notice that the structure of the deviation of the result from the exact-PQ one reported in Eq.~\eqref{eq:muon-coupling} can be obtained by combining all the previous considerations without the need of actually computing any loops.
As the symmetry breaking is realized by the simultaneous presence of four parameters, $Y_{V,V'},\,\Lambda,\,M_\psi$, one can parameterize the coupling of the ALP to muons as
\begin{equation}
     \sL_a\supset\frac{a}{f_a}\ov{\widehat{\mu}_L}i\gamma_5\widehat{\mu}_R\times m_\mu^\text{exp}\times\left(\frac{M_\psi(\alpha_1 Y_V+\alpha_2 Y_{V'})+\Lambda(\beta_1 Y_V+\beta_2 Y_{V'})}{M_\psi(\gamma_1 Y_V+\gamma_2 Y_{V'})+\Lambda(\delta_1 Y_V+\delta_2 Y_{V'})}\right)\,,
\end{equation}
where $\alpha_{1,2},\,\beta_{1,2},\,\gamma_{1,2},\,\delta_{1,2}$ are some unknown parameters that should be determined by the computation. By imposing the correction either to vanish or to be exactly $1$ when the symmetry is restored, depending on the charges of the fields, we can fix many of the constants. 
For example, when $Y_V=0$ we require the PQ-breaking correction to vanish as the PQ is vectorial for the muons at LO, meaning that $M_\psi \alpha_2 = - \Lambda \beta_2$.
Applying all conditions, it yields
\begin{equation}
    \sL_a\supset \frac{a}{f_a}\ov{\widehat{\mu}_L}i\gamma_5\widehat{\mu}_R\times m_\mu^\text{exp}\times\left(\frac{Y_V}{Y_V+\left(\frac{\gamma_2}{\delta_1}\right)\left(\frac{M_\psi}{\Lambda}\right) Y_{V'}}\right)\,.
\end{equation}
Such an expression correctly reproduces the result of the explicit computation of Eq.~\eqref{eq:muon-coupling} up to a numerical coefficient that cannot be determined solely by symmetry arguments.

%%%%%%%%%%%%%%%%%%%%%%%%%%%%%%%%%%%%%%%%%%%%%%%%%%%%%%%%%%%%%%%%%%%
%%%%%%%%%%%%%%%%%%%%%%%%%  Bibliography    %%%%%%%%%%%%%%%%%%%%%%%%
%%%%%%%%%%%%%%%%%%%%%%%%%%%%%%%%%%%%%%%%%%%%%%%%%%%%%%%%%%%%%%%%%%%
%
\footnotesize
\bibliography{bibliography}{}
\bibliographystyle{BiblioStyle}

\end{document}